\documentstyle[aps,amssymb]{revtex}

\begin{document}
\title{Glueball spectrum based on a rigorous three-dimensional relativistic
equation for two-gluon bound states II: calculation of the glueball spectrum}
\author{Jian-Xing Chen$^1$ and Jun-Chen Su$^{2*,3}$}
\address{1. Department of Physics, Peking University, Beijing 100871,\\
People's Republic of China\\
2. Department of Physics, Harbin Institute of Technology, Harbin 150006,\\
People's Republic of China\\
3. Center for Theoretical Physics, College of Physics, Jilin University,\\
Changchun 130023, People's Republic of China}
\date{}
\maketitle

\begin{abstract}
In the preceding paper, a rigorous three-dimensional\ relativistic equation
for two-gluon bound states was\ derived from the QCD with massive gluons and
represented in the angular momentum representation. In order to apply this
equation to calculate the glueball \ spectrum, in this paper,\ the equation
is\ recast in an equivalent three-dimensional relativistic equation
satisfied by the two-gluon positive energy state amplitude. The interaction
Hamiltonian in the equation is exactly derived and expressed as a
perturbative series. The first term in the series describes the one-gluon
exchange interaction which includes fully the retardation effect in it. This
term plus the linear confining potential are chosen to be the interaction
Hamiltonian\ and employed in the practical calculation. With the integrals
containing three and four spherical Bessel functions in the QCD vertices
being analytically calculated, the interaction Hamiltonian is given an
explicit expression in the angular momentum representation. Numerically
solving the relativistic equation with taking the contributions arising from
the retardation effect and the longitudinal mode of gluon fields into
account, a set of masses for the $0^{++},0^{-+},1^{++},1^{-+},2^{++}$ and $%
2^{-+\text{ }}$ glueball states are obtained and\ are in fairly good
agreement with the predictions given by the lattice simulation In addition,
some new glueball states are predicted.


*Corresponding author. E-mail address: junchens@public. cc.jl.cn
\end{abstract}

\section{Introduction}

As mentioned in the preceding paper, searching for glueballs, nowadays, is
an challenging task in particle physics.\ Since there are numerous technical
difficulties of giving unambiguous identifications of the glueballs in
experiment [1-15], it is expected that the existence of glueballs and their
properties could be precisely predicted by theoretical investigations so as
to guide the experimental searches. Various methods were proposed in the
past to serve such investigations [16-31]. However, the predictions given by
different methods are not consistent sometimes and even contradictory with
each other [32-33]. Of these methods, the lattice simulation [26-31] is\
considered to be faithful.\ Nevertheless, even for this method, there still
are controversies on the results given by different calculations.\ Apart
from the lattice computation, the nonrelativistic potential model [16-18],
the relativistic Dirac equation [22] and the Bethe-Salpeter (B-S) equation
[23-25] have recently been applied to evaluate the glueball spectrum. In
these methods, the interaction between gluons\ is constructed by two parts:
the short-range\ part which is described mainly by the one-gluon exchange
interaction and the long-range part which is represented by a
phenomenological confining potential. In the nonrelativistic potential
model, the short-range interaction is simulated by a potential which is
derived in the approximation of order $v^2/c^2$ where $v$ is the gluon
velocity and $c$ the light velocity with the assumption that the gluons in\
a glueball move not too fast. In a recent work by using this model [18],
with the choice of the lightest glueball masses given in the lattice
computation [29] as input, the authors obtained a series of two-gluon
glueball states with masses below $3GeV$. Except for some glueball masses\
which are in pretty good agreement with the lattice predictions, the other
calculated masses are apparently different from the lattice results. As was
emphasized in Ref.[18], to gain a physical solution to the lightest scalar
glueball, it is necessary to additionally introduce a phenomenological
smearing function to replace the $\delta $-function\ in the attractive
contact terms of the potential.\ Otherwise, the Hamiltonian would be
unbounded from below. This probably is\ an unnatural feature of the
nonrelativistic potential model. In Ref.[22], the calculation of the
glueball spectrum was performed by employing the relativistic Dirac equation
and showed only three theoretical results for the lowest glueball states $%
0^{++},2^{++}$ and $3^{++}$ some of which are not in so good agreement with
those given by\ lattice investigations. In the calculation, a Fermi-Breit
potential (the t-channel one-gluon exchange potential) was inserted into the
Dirac equation with the assumption that the nature and the force between two
gluons are the same as between two quarks. It seems that this assumption
ignores the difference between the potential for quarks and the one for
gluons. In addition, it would be mentioned that the Fermi-Breit potential is
derived in the nonrelativistic approximation of order $v^2/c^2$. Therefore,
the calculation is not fully relativistic. The relativistic calculation of
the glueball spectrum was carried out in the framework of B-S equation
[24,25]. Owing to the difficulty of solving a relativistic equation, only a
few states were predicted in these calculations. It is noted that in all the
\ previous applications of the B-S equation, the four-dimensional equation \
was recast in a three-dimensional form in the instantaneous approximation in
which the retardation effect is completely neglected. Another point we would
like to note here is that in the aforementioned works, the gluons are all
viewed as massive. Each of such gluons in general has three degrees of
freedom of polarization. Correspondingly, a gluon field should includes
three independent spatial components: two transverse\ fields and one
longitudinal field in the three-dimensional space. In this sense, we can say
that the Coulomb gauge as taken in Ref.[25] is inappropriate for the massive
gluons because in this gauge the longitudinal mode of the field is
completely eliminated. Similarly, in Ref.[24], only the transverse gluons
are taken into account even though the temporal gauge adopted in the work
allows existence of the longitudinal gluons.

In this paper, we intend to investigate the glueball spectrum based on the
three-dimensional relativistic equation for two-gluon bound states which was
derived in our former paper in the angular momentum representation. This
equation is actually a coupled set of equations satisfied by the four B-S
amplitudes for a glueball state: one is related to the positive energy
states of two gluons, the other three are related to the two gluon states in
which there is at least one gluon in the negative energy state. In the next
section, we will derive from this coupled equations an equivalent equation
obeyed by only the B-S amplitude of the \ glueball state for which the two
gluons are in the positive energy states and give the effective interaction
Hamiltonian in the equation a complete form. Since we are unable to compute
all the terms in the Hamiltonian at present, we are limited ourself to work
in a semi-phenomenological model by which the interaction Hamiltonian in the
equation is given by the one-gluon exchange kernel plus the phenomenological
linear confining kernel as was usually done in the previous literature
[16-18,23-25]. The new aspects of this paper which distinguish from the
previous works are: (1) The calculation is fully relativistic and hence
includes the contribution arising from all the relativistic effects to the
glueball masses; (2) The retardation effect of the one-gluon exchange
interaction is completely taken into account; (3) Apart from the transverse
modes of the gluon fields, the contribution from the longitudinal mode of
the field to the glueball spectrum is appropriately considered; (4) The
renormalization effect is considered by the effective QCD coupling constant
which was derived in the one-loop approximation\ and in a mass-dependent
subtraction in our previous work [34]. This coupling constant is not only
suitable in the high energy domain, but also in the low energy regime; (5)
We work in the angular momentum representation. In the this representation,
the glueball states are easily constructed. In particular, with completing
the radial integrals containing three and four spherical Bessel functions,
the gluon vertices are given explicit and analytical expressions which
greatly facilitate the numerical task of solving the equation. The
theoretical results obtained in this calculation are in quite good agreement
with those given in the lattice study [29-31], In addition, some new
predictions are presented.

The remainder of this paper is organized as follows. In Section II, we will
derive the three-dimensional equation satisfied by the gluon positive energy
state B-S amplitude from the coupled equations derived in the preceding
paper. In Section III, the interaction Hamiltonian obtained in the tree
diagram approximation will be discussed and its explicit expression will be
given. Section IV serves to\ derive the expression of the linear-wise
potential which is used to simulate the gluon confinement and incorporated
in the glueball equation for numerical calculations. In the last section,
the calculated results will be presented and some discussions will be made.
In Appendices A and B, the analytical expressions of three-line and
four-line gluon vertices are derived respectively.

\section{The three-dimensional equation for the gluon positive energy state
amplitude}

The three-dimensional equations derived in the preceding paper (which will
be called paper I later on) are shown in the following

\begin{equation}
(E_n-\omega _\alpha -\omega _\beta )\chi _{\alpha \beta
}(n)=\sum\limits_{\rho \sigma }K(\alpha \beta ;\rho \sigma ;E_n)\chi _{\rho
\sigma }(n),  \eqnum{2.1}
\end{equation}
where $E$ is the total energy of a glueball state, $\omega _\alpha $ and $%
\omega _\beta $ represent the energies of free gluons 1 and 2 respectively, $%
\chi _{\alpha \beta }(n)$ stands for the B-S amplitude describing the
glueball state which is defined by 
\begin{equation}
\chi _{\alpha \beta }(n)=\langle 0^{+}|{\bf a}_\alpha {\bf a}_\beta
|n\rangle ,  \eqnum{2.2}
\end{equation}
and $K(\alpha \beta ;\rho \sigma ;E)$ designates the interaction kernel
whose closed expression was derived in paper I. In the matrix notation, it
is of the form 
\begin{eqnarray}
K &=&\sum\limits_{i,j=1}^3\Lambda _j^{(i)}  \nonumber \\
\
&=&\{\sum\limits_{i=1}^3g_iS_i-\sum\limits_{i,j=1}^3g_iG_{ij}g_j+\sum%
\limits_{i,j=1}^3g_iG_iG^{-1}G_jg_j\}S^{-1},  \eqnum{2.3}
\end{eqnarray}
where the matrices in the above expression were clearly defined in paper I.
Noticing the definitions of $\omega _\alpha $ and \ ${\bf a}_\alpha $ (see
Eqs.(4.9) and (5.16) in paper I), 
\begin{equation}
\omega _\alpha =\left\{ 
\begin{array}{c}
\omega (k)\text{ \ if }\xi _\alpha =1 \\ 
-\omega (k)\text{ \ if }\xi _\alpha =-1
\end{array}
\right.  \eqnum{2.4}
\end{equation}
and 
\begin{equation}
{\bf a}_\alpha (k)=\left\{ 
\begin{array}{c}
{\bf a}_\alpha (k)\text{ \ if }\xi _\alpha =1 \\ 
\text{ \ }{\bf a}_\alpha ^{+}(k)\text{ \ if }\xi _\alpha =-1\text{ }
\end{array}
\right. ,  \eqnum{2.5}
\end{equation}
where the subscript $\alpha $ on the right hand side (RHS) of Eq.(2.5) is
defined without including $\xi _\alpha \ $and\ hence ${\bf a}_\alpha (k)$
and ${\bf a}_\alpha ^{+}(k)$ represent the annihilation and creation
operators respectively, the equation in Eq.(2.1) may be separately written
as 
\begin{equation}
\begin{tabular}{l}
$\lbrack E_n-\omega (k_1)-\omega (k_2)]\chi _{\alpha ^{+}\beta
^{+}}(n)=\sum\limits_{\rho \sigma }K(\alpha ^{+}\beta ^{+};\rho \sigma
;E_n)\chi _{\rho \sigma }(n),$ \\ 
$\lbrack E_n-\omega (k_1)+\omega (k_2)]\chi _{\alpha ^{+}\beta
^{-}}(n)=\sum\limits_{\rho \sigma }K(\alpha ^{+}\beta ^{-};\rho \sigma
;E_n)\chi _{\rho \sigma }(n),$ \\ 
$\lbrack E_n+\omega (k_1)-\omega (k_2)]\chi _{\alpha ^{-}\beta
^{+}}(n)=\sum\limits_{\rho \sigma }K(\alpha ^{-}\beta ^{+};\rho \sigma
;E_n)\chi _{\rho \sigma }(n),$ \\ 
$\lbrack E_n+\omega (k_1)+\omega (k_2)]\chi _{\alpha ^{-}\beta
^{-}}(n)=\sum\limits_{\rho \sigma }K(\alpha ^{-}\beta ^{-};\rho \sigma
;E_n)\chi _{\rho \sigma }(n),$%
\end{tabular}
\eqnum{2.6}
\end{equation}
where the superscripts $\pm $ in $\alpha ^{\pm }$ and $\beta ^{\pm }$ denote 
${\xi }_\alpha ,{\xi }_\beta {=\pm 1,}$the indices $\rho $ and $\sigma $
still include the indices ${\xi }_\rho $ and ${\xi }_\sigma $ respectively
and 
\begin{equation}
\begin{tabular}{l}
$\chi _{\alpha ^{+}\beta ^{+}}(n)=\langle 0^{+}|{\bf a}_\alpha {\bf a}_\beta
|n\rangle ,\chi _{\alpha ^{+}\beta ^{-}}(n)=\langle 0^{+}|{\bf a}_\alpha 
{\bf a}_\beta ^{+}|n\rangle ,$ \\ 
$\chi _{\alpha ^{-}\beta ^{+}}(n)=\langle 0^{+}|{\bf a}_\alpha ^{+}{\bf a}%
_\beta |n\rangle ,\chi _{\alpha ^{-}\beta ^{-}}(n)=\langle 0^{+}|{\bf a}%
_\alpha ^{+}{\bf a}_\beta ^{+}|n\rangle .$%
\end{tabular}
\eqnum{2.7}
\end{equation}

Following the procedure described in Ref.[35] for fermion systems, the
coupled equations in Eq.(2.6) can be reduced to an equivalent equation
satisfied by the B-S amplitude $\chi _{\alpha ^{+}\beta ^{+}}(n)$ for the
glueball state in which each of gluons is in its positive energy state. For
later convenience of derivation, we define 
\begin{equation}
\Delta _{ab}(E)=E-a\omega (k_1)-b\omega (k_2),  \eqnum{2.8}
\end{equation}
where the subscript $n$ in $E_n$ has been suppressed, $a,b=\pm 1$, 
\begin{equation}
\begin{tabular}{l}
$\phi _{++}(\alpha \beta ;E)=\chi _{\alpha ^{+}\beta ^{+}}(n),\phi
_{+-}(\alpha \beta ;E)=\chi _{\alpha ^{+}\beta ^{-}}(n),$ \\ 
$\phi _{-+}(\alpha \beta ;E)=\chi _{\alpha ^{-}\beta ^{+}}(n),\phi
_{--}(\alpha \beta ;E)=\chi _{\alpha ^{-}\beta ^{-}}(n)$%
\end{tabular}
\eqnum{2.9}
\end{equation}
and 
\begin{equation}
K_{abcd}(\alpha \beta ;\rho \sigma ;E)=K(\alpha ^a\beta ^b;\rho ^c\sigma
^d;E),  \eqnum{2.10}
\end{equation}
in which the $\alpha ,\beta ,\rho ,\sigma $ are defined without including
the index $\xi $. With the definitions given in Eqs.(2.8)-(2.10), the
equations in Eq.(2.6) can compactly be written as 
\begin{equation}
\Delta _{ab}(E)\phi _{ab}(\alpha \beta ;E)=\sum\limits_{cd}\sum\limits_{\rho
\sigma }K_{abcd}(\alpha \beta ;\rho \sigma ;E)\phi _{cd}(\rho \sigma ;E), 
\eqnum{2.11}
\end{equation}
where $a,b,c,d=\pm 1.$ In the product space of momentum $k_{1,}$ $k_2$ and
angular momentum marked by $\alpha $, the above equations may be written in
the matrix form 
\begin{equation}
\Delta _{++}(E)\phi _{++}(E)=K_{++++}(E)\phi _{++}(E)+\sum\limits_{cd\neq
++}K_{++cd}(E)\phi _{cd}(E),  \eqnum{2.12}
\end{equation}
\begin{equation}
\Delta _{ab}(E)\phi _{ab}(E)=K_{ab++}(E)\phi _{++}(E)+\sum\limits_{cd\neq
++}K_{abcd}(E)\phi _{cd}(E),  \eqnum{2.13}
\end{equation}
where $ab\neq ++$ and the terms related to $\phi _{++}(E)$ have been
separated out from the others. Furthermore, In the space spanned by $\phi
_{ab}(E)$ with $ab\neq ++$, we use the matrix representation defined as
follows 
\begin{equation}
\begin{tabular}{l}
$\psi (E)=\phi _{++}(E),\phi (E)=\{\phi _{ab}(E)\},$ \\ 
$\Delta _{+}(E)=\Delta _{++}(E),\Delta (E)=\{\Delta _{ab}(E)\},$ \\ 
$K_{+}(E)=K_{++++}(E),\overline{K}^t(E)=\{K_{++cd}(E)\},$ \\ 
$\overline{G}(E)=\{K_{ab++}(E)/\Delta _{ab}(E)\},\},$ \\ 
$G(E)=\{K_{abcd}(E)/\Delta _{ab}(E)\}.$%
\end{tabular}
\eqnum{2.14}
\end{equation}
According to these definitions, Eqs.(2.12) and (2.13) may be written in the
full matrix form 
\begin{equation}
\Delta _{+}(E)\psi (E)=K_{+}(E)\psi (E)+\overline{K}^t(E)\phi (E), 
\eqnum{2.15}
\end{equation}
\begin{equation}
\phi (E)=\overline{G}(E)\psi (E)+G(E)\phi (E).  \eqnum{2.16}
\end{equation}
Solving the equation (2.16), we obtain 
\begin{equation}
\phi (E)=\frac 1{1-G(E)}\overline{G}(E)\psi (E).  \eqnum{2.17}
\end{equation}
Substituting the above expression into Eq.(2.15), we finally arrive at 
\begin{equation}
\Delta _{+}(E)\psi (E)=V(E)\psi (E),  \eqnum{2.18}
\end{equation}
where 
\begin{equation}
V(E)=K_{+}(E)+\overline{K}^t(E)\frac 1{1-G(E)}\overline{G}(E),  \eqnum{2.19}
\end{equation}
which is\ identified itself with the interaction Hamiltonian. Noticing the\
definition 
\begin{equation}
\frac 1{1-G(E)}=\sum\limits_{n=0}G^{(n)}(E).  \eqnum{2.20}
\end{equation}
Eq.(2.19) can be written as 
\begin{equation}
V(E)=\sum\limits_{n=0}V^{(n)}(E),  \eqnum{2.21}
\end{equation}
where 
\begin{equation}
\begin{tabular}{l}
$V^{(0)}(E)=K_{+}(E),$ \\ 
$V^{(1)}(E)=\overline{K}^t(E)\overline{G}(E),$ \\ 
$V^{(2)}(E)=\overline{K}^t(E)G(E)\overline{G}(E)$ \\ 
$\cdot \cdot \cdot \cdot \cdot \cdot \cdot .$%
\end{tabular}
\eqnum{2.22}
\end{equation}
According to the definitions in Eq.(2.14), Eqs.(2.21) and (2.22) may be
explicitly written as 
\begin{equation}
V(\alpha \beta ;\gamma \delta ;E)=\sum\limits_{n=0}V^{(n)}(\alpha \beta
;\gamma \delta ;E),  \eqnum{2.23}
\end{equation}
where 
\begin{equation}
V^{(0)}(\alpha \beta ;\gamma \delta ;E)=K(\alpha ^{+}\beta ^{+};\gamma
^{+}\delta ^{+};E),  \eqnum{2.24}
\end{equation}
\begin{equation}
V^{(1)}(\alpha \beta ;\gamma \delta ;E)=\sum\limits_{ab\neq
++}\sum\limits_{\rho \sigma }\frac{K(\alpha ^{+}\beta ^{+};\rho ^a\sigma
^b;E)K(\rho ^a\sigma ^b;\gamma ^{+}\delta ^{+}E)}{E-a\omega (k_1)-b\omega
(k_2)},  \eqnum{2.25}
\end{equation}
\begin{equation}
\begin{tabular}{l}
$V^{(2)}(\alpha \beta ;\gamma \delta ;E)$ \\ 
$=\sum\limits_{ab\neq ++}\sum\limits_{cd\neq ++}\sum\limits_{\rho \sigma
}\sum\limits_{\mu \nu }\frac{K(\alpha ^{+}\beta ^{+};\rho ^a\sigma
^b;E)K(\rho ^a\sigma ^b;\mu ^c\nu ^d;E)K(\mu ^c\nu ^d;\gamma ^{+}\delta
^{+};E)}{(E-a\omega (k_1)-b\omega (k_2))(E-c\omega (k_1)-d\omega (k_2))}$ \\ 
$\cdot \cdot \cdot \cdot \cdot \cdot ,$%
\end{tabular}
\eqnum{2.26}
\end{equation}
where we have used the notation shown in Eqs.(2.8) and (2.10). From
Eqs.(2.19)-(2.26), \ it is clear to see that the negative energy states of
two gluons act as intermediate states to appear in the effective interaction
Hamiltonian and give contributions to the higher order terms in the
Hamiltonian. The equation (2.18)\ written in an explicit form\ is such that 
\begin{equation}
\lbrack E-\omega (k_1)-\omega (k_2)]\psi (\alpha \beta
;E)=\sum\limits_{\gamma \delta }V(\alpha \beta ;\gamma \delta ;E)\psi
(\gamma \delta ;E).  \eqnum{2.27}
\end{equation}
This is just the equation satisfied by the gluon positive energy state
amplitudes for the glueball states in which 
\begin{equation}
\psi (\alpha \beta ;E)=\chi _{\alpha ^{+}\beta ^{+}}(n).  \eqnum{2.28}
\end{equation}
\ 

\section{The Hamiltonian given in the lowest order approximation}

In this section, we plan to discuss the interaction Hamiltonian\ in the tree
diagram approximation of the order of $g^2$\ here\ $g$ is the QCD coupling
constant. This Hamiltonian can only be given by the term shown in Eq.(2.24)
because the other terms in the effective Hamiltonian give the contributions
which are all higher than $g^2$. In general, the $K(\alpha ^{+}\beta
^{+};\gamma ^{+}\delta ^{+};E)$ should be calculated according to the
expression denoted in Eq.(2.3) which includes three parts. The last part in
Eq.(2.3) plays the role of cancelling the B-S reducible diagrams contained
in the first two parts and gives no contribution of the order $g^2$. This is
because (1) the coefficients $g_1(\alpha \beta ;\rho \sigma \lambda )$ and $%
g_3(\alpha \beta ;\rho \sigma \lambda )$ are proportional to $g$ and $%
g_2(\alpha \beta ;\rho \sigma \tau \lambda )$ is proportional to $g^2$; (2)
the Green's functions $G_1(\rho \sigma \lambda ;\gamma \delta ;t_1-t_2)$ \
and $G_3(\rho \sigma \lambda ;\gamma \delta ;t_1-t_2)$ vanish in the lowest
order approximation. As for the first part in Eq.(2.3), in the lowest order
approximation, it is easy to verify that 
\begin{equation}
\begin{tabular}{l}
$S_1(\rho \sigma \lambda ,\gamma \delta )=\langle 0|[:{\bf a}_\rho {\bf a}%
_\sigma :{\bf a}_\lambda ,{\bf a}_\gamma {\bf a}_\delta ]|0\rangle =0,$ \\ 
$S_2(\rho \sigma \tau \lambda ,\gamma \delta )=\langle 0|[:{\bf a}_\rho {\bf %
a}_\sigma {\bf a}_\tau :{\bf a}_\lambda ,{\bf a}_\gamma {\bf a}_\delta
]|0\rangle =0,$ \\ 
$S_3(\rho \sigma \lambda ,\gamma \delta )=\langle 0|[:{\bf c}_\rho ^{+}{\bf c%
}_\sigma :{\bf a}_\lambda ,{\bf a}_\gamma {\bf a}_\delta ]|0\rangle =0,$%
\end{tabular}
\eqnum{3.1}
\end{equation}
where $|0\rangle $ denotes the bare vacuum state. Therefore, we only need to
consider the second part in Eq.(2.3). In this part, the term related to $%
g_2(\alpha \beta ;\rho \sigma \tau \lambda )$ gives the contribution of
order $g^4$ in the lowest approximation and hence is beyond our
consideration. For the terms associated with $g_3(\alpha \beta ;\rho \sigma
\lambda ),$the relevant Green's functions vanish in the lowest order
approximation. Thus, we are only left with terms in the second part of
Eq.(2.3) such that 
\begin{equation}
K^0(\alpha ^{+}\beta ^{+};\gamma ^{+}\delta ^{+};E)=\sum\limits_{\rho \sigma
}\Lambda (\alpha ^{+}\beta ^{+};\rho \sigma ;E)S^{-1}(\rho \sigma ;\gamma
^{+}\delta ^{+}),  \eqnum{3.2}
\end{equation}
where 
\begin{equation}
\begin{tabular}{l}
$\Lambda (\alpha ^{+}\beta ^{+};\rho \sigma ;E)$ \\ 
$=-\sum\limits_{\xi \eta \lambda }\sum\limits_{\mu \nu \tau }g_1(\alpha
^{+}\beta ^{+};\xi \eta \lambda )G_{11}(\xi \eta \lambda ;\mu \nu \tau
;E)g_1(\mu \nu \tau ;\rho \sigma )$%
\end{tabular}
\eqnum{3.3}
\end{equation}
and the indices $\rho ,\sigma ,\cdot \cdot \cdot $ should be understood as $%
\rho ^{\pm },\sigma ^{\pm },\cdot \cdot \cdot .$

Let us first compute the inverse $S^{-1}(\rho \sigma ;\gamma ^{+}\delta
^{+}) $. For this purpose, we operate on the both sides of Eq.(3.2) with $S$
\ from the right and get 
\begin{equation}
\Lambda (\alpha ^{+}\beta ^{+};\rho \sigma ;E)=\sum\limits_{\gamma \delta
}K^0(\alpha ^{+}\beta ^{+};\gamma ^{+}\delta ^{+};E)S(\gamma ^{+}\delta
^{+};\rho \sigma ).  \eqnum{3.4}
\end{equation}
It is easy to verify that except for $S(\gamma ^{+}\delta ^{+};\rho
^{-}\sigma ^{-}),$ the $S(\gamma ^{+}\delta ^{+};\rho ^{-}\sigma ^{+}),$ $%
S(\gamma ^{+}\delta ^{+};\rho ^{+}\sigma ^{-})$ and $S(\gamma ^{+}\delta
^{+};\rho ^{+}\sigma ^{+})$ are all vanishing in the lowest order
approximation. As for the $S(\gamma ^{+}\delta ^{+};\rho ^{-}\sigma ^{-})$,
we have 
\begin{equation}
S(\gamma ^{+}\delta ^{+};\rho ^{-}\sigma ^{-})=\delta _{\gamma \rho }\delta
_{\delta \sigma }+\delta _{\gamma \sigma }\delta _{\delta \rho }. 
\eqnum{3.5}
\end{equation}
Substituting Eq.(3.5) in Eq.(3.4), we find 
\begin{equation}
\Lambda (\alpha ^{+}\beta ^{+};\rho ^{-}\sigma ^{-};E)=K^0(\alpha ^{+}\beta
^{+};\rho ^{+}\sigma ^{+};E)+K^0(\alpha ^{+}\beta ^{+};\sigma ^{+}\rho
^{+};E).  \eqnum{3.6}
\end{equation}
Since we may interchange the indices $\rho $ and $\sigma $ in Eq.(2.1) or in
Eq.(2.27), noticing $\chi _{\rho \sigma }(n)=\chi _{\sigma \rho }(n)$ or $%
\psi (\rho \sigma ;E)=\psi (\sigma \rho ;E)$, we can write from Eq.(3.6) the
following relation 
\begin{equation}
K^0(\alpha ^{+}\beta ^{+};\rho ^{+}\sigma ^{+};E)=\frac 12\Lambda (\alpha
^{+}\beta ^{+};\rho ^{-}\sigma ^{-};E),  \eqnum{3.7}
\end{equation}
which means that we may set 
\begin{equation}
S^{-1}(\rho \sigma ;\gamma ^{+}\delta ^{+})=S^{-1}(\rho ^{-}\sigma
^{-};\gamma ^{+}\delta ^{+})=\frac 12\delta _{\gamma \rho }\delta _{\delta
\sigma }.  \eqnum{3.8}
\end{equation}
Combining Eqs.(3.3) and (3.7), we have 
\begin{equation}
\begin{tabular}{l}
$K^0(\alpha ^{+}\beta ^{+};\rho ^{+}\sigma ^{+};E)$ \\ 
$=-\frac 12\sum\limits_{\gamma \delta \lambda }\sum\limits_{\mu \nu \tau
}g_1(\alpha ^{+}\beta ^{+};\gamma \delta \lambda )G_{11}(\gamma \delta
\lambda ;\mu \nu \tau ;E)g_1(\mu \nu \tau ;\rho ^{-}\sigma ^{-}).$%
\end{tabular}
\eqnum{3.9}
\end{equation}
In accordance with the definition of $g_1(\alpha \beta ;\gamma \delta
\lambda )$ and $g_1(\mu \nu \tau ;\rho \sigma )$ (see Eqs.(5.7), (5.9),
(5.10) and (6.4) in paper I), we can write 
\begin{equation}
\begin{tabular}{l}
$g_1(\alpha ^{+}\beta ^{+};\gamma \delta \lambda )=\sum\limits_\tau
f_1(\gamma \delta \tau )\Delta _{\alpha ^{+}\beta ^{+};\tau \lambda },$ \\ 
$g_1(\mu \nu \tau ;\rho ^{-}\sigma ^{-})=\sum\limits_\lambda f_1(\mu \nu
\lambda )\Delta _{\rho ^{-}\sigma ^{-};\lambda \tau },$%
\end{tabular}
\eqnum{3.10}
\end{equation}
where 
\begin{equation}
\begin{tabular}{l}
$\Delta _{\alpha ^{+}\beta ^{+};\tau \lambda }=\Delta _{\alpha ^{+}\tau
}\delta _{\beta ^{+}\lambda }+\Delta _{\beta ^{+}\tau }\delta _{\alpha
^{+}\lambda },$ \\ 
$\Delta _{\rho ^{-}\sigma ^{-};\lambda \tau }=\Delta _{\rho ^{-}\lambda
}\delta _{\sigma ^{-}\tau }+\Delta _{\sigma ^{-}\lambda }\delta _{\rho
^{-}\tau }$%
\end{tabular}
\eqnum{3.11}
\end{equation}
and 
\begin{equation}
f_1(\alpha \beta \gamma )=A(\alpha \beta \gamma )+A(\alpha \gamma \beta
)+A(\gamma \alpha \beta ),  \eqnum{3.12}
\end{equation}
here $A(\alpha \beta \gamma )$ is the three-line gluon vertex given in the
angular momentum representation. Considering the expressions in Eq.(3.11)
and the fact that only $\Delta _{\alpha ^{+}\beta ^{-}}=-\Delta _{\alpha
^{-}\beta ^{+}}=\delta _{\alpha \beta }$ are nonvanishing for $\Delta
_{\alpha \beta }$, Eq.(3.10)\ can be represented as 
\begin{equation}
\begin{array}{l}
g_1(\alpha ^{+}\beta ^{+};\gamma \delta \lambda )=f_1(\gamma \delta \alpha
^{-})\delta _{\beta ^{+}\lambda }+f_1(\gamma \delta \beta ^{-})\delta
_{\alpha ^{+}\lambda }, \\ 
g_1(\mu \nu \tau ;\rho ^{-}\sigma ^{-})=-f_1(\mu \nu \rho ^{+})\delta
_{\sigma ^{-}\tau }-f_1(\mu \nu \sigma ^{+})\delta _{\rho ^{-}\tau }.
\end{array}
\eqnum{3.13}
\end{equation}
On inserting Eq.(3.13) into Eq.(3.9), one gets 
\begin{equation}
\begin{tabular}{l}
$K^0(\alpha ^{+}\beta ^{+};\rho ^{+}\sigma ^{+};E)=\frac 12[\Lambda (\alpha
^{+}\beta ^{+};\rho ^{+}\sigma ^{+};E)+(\alpha \leftrightarrow \beta )$ \\ 
$+(\rho \leftrightarrow \sigma )+(\alpha \leftrightarrow \beta ,\rho
\leftrightarrow \sigma )],$%
\end{tabular}
\eqnum{3.14}
\end{equation}
where 
\begin{equation}
\Lambda (\alpha ^{+}\beta ^{+};\rho ^{+}\sigma ^{+};E)=\sum\limits_{\gamma
\delta ,\mu \nu }f_1(\gamma \delta \alpha ^{-})G_{11}(\gamma \delta \beta
^{+};\mu \nu \sigma ^{-};E)f_1(\mu \nu \rho ^{+})  \eqnum{3.15}
\end{equation}
and the other terms in Eq.(3.14) can be obtained from the first term by
exchanging the indices\ as shown in Eq.(3.14). Now, let us calculate the
Green's functions $G_{11}(\gamma \delta \beta ^{+};\mu \nu \sigma ^{-};E)$
in the lowest order approximation which are the Fourier transform of the
Green functions $G_{11}(\gamma \delta \beta ^{+};\mu \nu \sigma
^{-};t_1-t_2) $. With the aid of Wick theorem, it can be found that only the
following Green's function is nonvanishing 
\begin{equation}
\begin{tabular}{l}
$G_{11}(\gamma ^{+}\delta ^{+}\beta ^{+};\mu ^{-}\nu ^{-}\sigma
^{-};t_1-t_2) $ \\ 
$=\langle 0|T\{:{\bf a}_\gamma (t_1){\bf a}_\delta (t_1):{\bf a}_\beta (t_1):%
{\bf a}_\mu ^{+}(t_2){\bf a}_\nu ^{+}(t_2):{\bf a}_\sigma
^{+}(t_2)\}|0\rangle ,$%
\end{tabular}
\eqnum{3.16}
\end{equation}
where ${\bf a}_\gamma (t_1)$ and ${\bf a}_\sigma ^{+}(t_2)$ are the
annihilation and creation operators in the interaction picture. Noticing 
\begin{equation}
{\bf a}_\gamma (t_1)={\bf a}_\gamma e^{-i\omega _\gamma t_1},{\bf a}_\sigma
^{+}(t_2)={\bf a}_\sigma ^{+}e^{i\omega _\sigma t_2}  \eqnum{3.17}
\end{equation}
and applying the Wick theorem, we find

\begin{equation}
\begin{array}{l}
G_{11}(\gamma ^{+}\delta ^{+}\beta ^{+};\mu ^{-}\nu ^{-}\sigma ^{-};t_1-t_2)
\\ 
=\theta (t_1-t_2)e^{-i(\omega _\gamma +\omega _\delta +\omega _\beta
)(t_1-t_2)}[\delta _{\gamma \mu }\delta _{\delta \nu }\delta _{\beta \sigma
}+\delta _{\gamma \nu }\delta _{\delta \mu }\delta _{\beta \sigma } \\ 
+\delta _{\gamma \sigma }\delta _{\delta \mu }\delta _{\beta \nu }+\delta
_{\gamma \sigma }\delta _{\delta \nu }\delta _{\beta \mu }+\delta _{\gamma
\mu }\delta _{\delta \sigma }\delta _{\beta \nu }+\delta _{\gamma \nu
}\delta _{\delta \sigma }\delta _{\beta \mu }].
\end{array}
\eqnum{3.18}
\end{equation}
By the Fourier transformation and using the familiar integral representation
of the step function, we obtain 
\begin{equation}
\begin{array}{l}
G_{11}(\gamma ^{+}\delta ^{+}\beta ^{+};\mu ^{-}\nu ^{-}\sigma ^{-};E) \\ 
=\frac 1i\int_{-\infty }^{+\infty }dte^{iEt}G_{11}(\gamma ^{+}\delta
^{+}\beta ^{+};\mu ^{-}\nu ^{-}\sigma ^{-};t) \\ 
=\frac 1{E-\omega _\gamma -\omega _\delta -\omega _\beta +i\epsilon }[\delta
_{\gamma \mu }\delta _{\delta \nu }\delta _{\beta \sigma }+\delta _{\gamma
\nu }\delta _{\delta \mu }\delta _{\beta \sigma }+\delta _{\gamma \sigma
}\delta _{\delta \mu }\delta _{\beta \nu } \\ 
+\delta _{\gamma \sigma }\delta _{\delta \nu }\delta _{\beta \mu }+\delta
_{\gamma \mu }\delta _{\delta \sigma }\delta _{\beta \nu }+\delta _{\gamma
\nu }\delta _{\delta \sigma }\delta _{\beta \mu }].
\end{array}
\eqnum{3.19}
\end{equation}
Substituting the above expression in\ Eq.(3.15), we are led to 
\begin{equation}
\begin{array}{l}
\Lambda (\alpha ^{+}\beta ^{+};\rho ^{+}\sigma ^{+};E) \\ 
=\sum\limits_{\gamma \delta }\frac 1{E_n-\omega _\gamma -\omega _\delta
-\omega _\beta +i\epsilon }f_1(\gamma ^{+}\delta ^{+}\alpha
^{-})\{[f_1(\gamma ^{-}\delta ^{-}\rho ^{+}) \\ 
+f_1(\delta ^{-}\gamma ^{-}\rho ^{+})]\delta _{\beta \sigma }+[f_1(\delta
^{-}\beta ^{-}\rho ^{+})+f_1(\beta ^{-}\delta ^{-}\rho ^{+})]\delta _{\gamma
\sigma } \\ 
+[f_1(\gamma ^{-}\beta ^{-}\rho ^{+})+f_1(\beta ^{-}\gamma ^{-}\rho
^{+})]\delta _{\delta \sigma }\}.
\end{array}
\eqnum{3.20}
\end{equation}
Observing the above expression\ of $\Lambda (\alpha ^{+}\beta ^{+};\rho
^{+}\sigma ^{+};E)$, we see, the first term containing $\delta _{\beta
\sigma }$\ is an unconnected term. It gives the one-loop correction to the
gluon propagator whose effect will be included in the QCD effective coupling
constant by the renormalization procedure. The second term proportional to $%
\delta _{\gamma \sigma }$ describes the one-gluon exchange interaction
between two gluons. The third term is the exchanged term for the one-gluon
exchange interaction. By dropping the first term, we have 
\begin{equation}
\begin{tabular}{l}
$\Lambda (\alpha ^{+}\beta ^{+};\rho ^{+}\sigma ^{+};E)$ \\ 
$=\sum\limits_\tau \frac{[f_1(\tau ^{+}\sigma ^{+}\alpha ^{-})+f_1(\sigma
^{+}\tau ^{+}\alpha ^{-})][f_1(\tau ^{-}\beta ^{-}\rho ^{+})+f_1(\beta
^{-}\tau ^{-}\rho ^{+})]}{E-\omega _\tau -\omega _\sigma -\omega _\beta
+i\epsilon },$%
\end{tabular}
\eqnum{3.21}
\end{equation}
where the summation over $\tau $ is performed with respect to the gluon
intermediate states and the function $f_1$ was represented in Eq.(3.12)\ in
terms of the function $A(\alpha \beta \gamma )\ $whose explicit expression
is derived in Appendix A and shown in the following. 
\begin{equation}
\begin{array}{l}
A(\alpha _1\alpha _2\alpha _3)=-\frac g2(\frac 2\pi )^{\frac 32%
}f^{abc}k_3\prod\limits_{i=1}^3\frac 1{\sqrt{2\omega (k_i)}}k_iB^{\xi
_i}(l_i)_{\lambda _i\tau _i}T_{\lambda _3\lambda _j} \\ 
\times J_{l_1^{^{\prime }}l_2^{^{\prime }}l_3^{^{\prime
}}}(k_1,k_2,k_3)\Gamma (l_i,l_i^{\prime },m_i,\xi _i),
\end{array}
\eqnum{3.22}
\end{equation}
where 
\begin{equation}
\begin{array}{l}
J_{l_1^{\prime }l_2^{\prime }l_3^{^{\prime }}}(k_1,k_2,k_3)=\frac{\pi ^{%
\frac 52}}4(-1)^{\frac 12(l_1^{\prime }+l_2^{\prime }+l_3^{^{\prime }})} \\ 
\times \sum\limits_{\mu _1,\mu _2,\mu _3=0}^\infty \delta _{2(\mu _1+\mu
_2+\mu _3),l_1^{\prime }+l_2^{\prime }+l_3^{\prime }}\prod\limits_{i=1}^3%
\frac{k_i^{2\mu _i-l_i^{^{\prime }}-1}}{\Gamma (\mu _i+1)\Gamma (\mu
_i-l_i^{\prime }+\frac 12)},
\end{array}
\eqnum{3.23}
\end{equation}
\begin{equation}
\begin{tabular}{l}
$\Gamma (l_i,l_i^{\prime },m_i,\eta _i)$ \\ 
$=\frac i{\sqrt{2\pi }}\prod\limits_{i=1}^3(-1)^{(l_i+l_i^{\prime
}+m_i+1)\sin [\frac{(1-\eta _i)\pi }4]}[(2l_i+1)(2l_i^{\prime }+1)]^{\frac 12%
}$ \\ 
$\times \left( 
\begin{array}{ccc}
l_1^{\prime } & l_2^{\prime } & l_3^{\prime } \\ 
0 & 0 & 0
\end{array}
\right) \left( 
\begin{array}{ccc}
l_1 & l_2 & l_3 \\ 
m_1 & m_2 & m_3
\end{array}
\right) \left\{ 
\begin{array}{ccc}
1 & 1 & 1 \\ 
l_1^{\prime } & l_2^{\prime } & l_3^{\prime } \\ 
l_1 & l_2 & l_3
\end{array}
\right\} ,$%
\end{tabular}
\eqnum{3.24}
\end{equation}
\ $T_{\lambda _3\lambda _j}$ and\ $B^{\xi _i}(l_i)_{\lambda _i\tau _i}$ are
defined respectively in (A.5) and (A.15). It is noted that for a given set
of angular momenta, due to the restriction of $\delta _{2(\mu _1+\mu _2+\mu
_3),\allowbreak \allowbreak l_1^{\prime }+l_2^{\prime }+l_3^{\prime }}$,
only a few terms in the series of Eq.(3.23) survive.

\section{The glueball equation with inclusion of the confinement}

As shown in Sec.II, the interaction Hamiltonian in the exact relativistic
equation for the glueball states is much complicated. In practical
investigations, usually, one only considers the lowest order term in the
Hamiltonian which was explicitly derived in the angular momentum
representation in the preceding section. Obviously, in order to get
reasonable theoretical results, it is necessary to introduce a certain
confining potential to phenomenologically simulate all the higher order
terms in the Hamiltonian [16-18, 23-25]. How to determine the form of
confining interaction in the relativistic equation ? In this paper, as was
similarly done for the quark-antiquark system [36], we introduce the
confining Hamiltonian operator in such a manner 
\begin{eqnarray}
&&\ H_c  \nonumber \\
\ &=&\frac 12\int d^3x_1d^3x_2f^{abe}f^{cde}\overrightarrow{A^a}(%
\overrightarrow{x}_1)\cdot \overrightarrow{A^b}(\overrightarrow{x}_1)V(|%
\overrightarrow{x}_1-\overrightarrow{x}_2|)\overrightarrow{A^c}(%
\overrightarrow{x}_2)\cdot \overrightarrow{A^d}(\overrightarrow{x}_2), 
\eqnum{4.1}
\end{eqnarray}
where $\overrightarrow{A^a}(\overrightarrow{x})$ stands for the gluon field
operator and $V(|\overrightarrow{x}_1-\overrightarrow{x}_2|)$ the confining
potential which will be taken to be the linear one [37] 
\begin{equation}
V(|\overrightarrow{x}_1-\overrightarrow{x}_2|)=\gamma |\overrightarrow{x}_1-%
\overrightarrow{x}_2|,  \eqnum{4.2}
\end{equation}
here the parameter $\gamma $ designates the strength of the confining
potential. When the gluon field operators in Eq.(4.1) are replaced by their
expansions in terms of the multipole fields (see the expansion given in
Eq.(4.14) in paper I), Eq.(4.1) will be represented as 
\begin{equation}
H_c=\sum\limits_{\alpha _1\alpha _2\alpha _3\alpha _4}V_c(\alpha _1\alpha
_3;\alpha _2\alpha _4):{\bf a}_{\alpha _1}{\bf a}_{\alpha _2}{\bf a}_{\alpha
_3}{\bf a}_{\alpha _4}:,  \eqnum{4.3}
\end{equation}
where 
\begin{equation}
\begin{tabular}{l}
$V_c(\alpha _1\alpha _3;\alpha _2\alpha _4)$ \\ 
$=\frac 12f^{abe}f^{cde}\int d^3x_1d^3x_2\overrightarrow{A}_{\beta
_1}^{\lambda _1}(\overrightarrow{x}_1)\cdot \overrightarrow{A}_{\beta
_2}^{\lambda _2}(\overrightarrow{x}_1)V(|\overrightarrow{x}_1-%
\overrightarrow{x}_2|)\overrightarrow{A}_{\beta _3}^{\lambda _3}(%
\overrightarrow{x}_2)\cdot \overrightarrow{A}_{\beta _4}^{\lambda _4}(%
\overrightarrow{x}_2).$%
\end{tabular}
\eqnum{4.4}
\end{equation}
Here the symbols $\lambda _i$ and $\beta _i$ \ were defined in Appendix A.
This is just the wanted confining potential written in the angular momentum
representation which will be inserted into the relativistic equation. By
using the Fourier transformation 
\begin{equation}
|\overrightarrow{x}_1-\overrightarrow{x}_2|=-\int \frac{d^3q}{(2\pi )^3}%
\frac{8\pi }{\overrightarrow{q}^4}e^{i\overrightarrow{q}\cdot (%
\overrightarrow{x}_1-\overrightarrow{x}_2)},  \eqnum{4.5}
\end{equation}
the expansion for the plane wave function 
\begin{equation}
e^{\overrightarrow{p}\cdot \overrightarrow{x}}=4\pi
\sum\limits_{lm}i^lj_l(pr)Y_{lm}^{*}(\widehat{p})Y_{lm}(\widehat{x}) 
\eqnum{4.6}
\end{equation}
and the expression for the scalar multipole field 
\begin{equation}
A_{lm}^S(k\overrightarrow{x})=\sqrt{\frac 2\pi }kj_l(kr)Y_{lm}(\widehat{x}),
\eqnum{4.7}
\end{equation}
we can write 
\begin{equation}
V(|\overrightarrow{x}_1-\overrightarrow{x}_2|)=-8\pi \gamma
\sum\limits_{lm}\int_0^\infty dq\frac 1{q^4}A_{lm}^S(q\overrightarrow{x}%
_1)A_{lm}^{S^{*}}(q\overrightarrow{x}_2).  \eqnum{4.8}
\end{equation}
Substitution of this expression into Eq.(4.4) leads to 
\begin{equation}
V_c(\alpha _1\alpha _3;\alpha _2\alpha _4)=-4\pi \gamma
f^{abe}f^{cde}\sum\limits_{lm}\int_0^\infty dq\frac 1{q^4}D_{\alpha _1\alpha
_2\tau }D_{\alpha _3\alpha _4\tau }^{*},  \eqnum{4.9}
\end{equation}
where 
\begin{equation}
\begin{array}{c}
D_{\alpha _1\alpha _2\tau }=\int d^3x_1\overrightarrow{A}_{\beta
_1}^{\lambda _1}(\overrightarrow{x}_1)\cdot \overrightarrow{A}_{\beta
_2}^{\lambda _2}(\overrightarrow{x}_1)A_\tau ^S(\overrightarrow{x}_1), \\ 
D_{\alpha _3\alpha _4\tau }=\int d^3x_2\overrightarrow{A}_{\beta
_3}^{\lambda _3}(\overrightarrow{x}_2)\cdot \overrightarrow{A}_{\beta
_4}^{\lambda _4}(\overrightarrow{x}_2)A_\tau ^{S^{*}}(\overrightarrow{x}_2),
\end{array}
\eqnum{4.10}
\end{equation}
here $\tau =(q,l,m)$.

Completely analogous to the calculations described in\ Appendix A, it is
easy to derive the following expression 
\begin{equation}
\begin{array}{l}
V_c(\alpha _1\alpha _3;\alpha _2\alpha _4) \\ 
=-\frac{32}{\pi ^2}\lambda f^{abe}f^{cde}\sum\limits_{lm}\int_0^\infty dq%
\frac 1{q^2}\prod\limits_{i=1}^4k_iB^{\xi _i}(l_i)_{\lambda _i\tau _i} \\ 
\times J_{l_1^{\prime }l_2^{\prime }l}(k_1,k_2,q)J_{l_3^{\prime }l_4^{\prime
}l}(k_3,k_4,q) \\ 
\times \widehat{\Gamma }(l_i,l_i^{\prime },m_i,\eta _i,l,m,1)\widehat{\Gamma 
}(l_j,l_j^{\prime },m_j,\eta _j,l,m,-1),
\end{array}
\eqnum{4.11}
\end{equation}
where $i=1,2$ $,$ $j=3,4$ , 
\begin{equation}
\begin{array}{l}
\widehat{\Gamma }(l_i,l_i^{\prime },m_i,\eta _i,l,m,\eta ) \\ 
\equiv \int d\Omega (\widehat{x})\overrightarrow{Y}_{l_1l_1^{\prime
}m_1}^{\eta _1}(\widehat{x})\cdot \overrightarrow{Y}_{l_2l_2^{\prime
}m_2}^{\eta _2}(\widehat{x})Y_{lm}^\eta (\widehat{x}) \\ 
=\prod\limits_{i=1}^2(-1)^{(l_i+l_i^{\prime }+m_i+1)\sin [\frac{(1-\eta
_i)\pi }4]}(-1)^{m\sin [\frac{(1-\eta )\pi }4]} \\ 
\times \widetilde{\Gamma }(l_i,l_i^{\prime },\eta _im_i,l,\eta m)
\end{array}
\eqnum{4.12}
\end{equation}
with 
\begin{equation}
\begin{array}{l}
\widetilde{\Gamma }(l_i,l_i^{\prime },m_i,l,m)\equiv \int d\Omega (\widehat{x%
})\overrightarrow{Y}_{l_1l_1^{\prime }m_1}(\widehat{x})\cdot \overrightarrow{%
Y}_{l_2l_2^{\prime }m_2}(\widehat{x})Y_{lm}(\widehat{x}) \\ 
=(-1)^{l_1+l_2^{\prime }+l}[\frac{(2l_1+1)(2l_2+1)(2l_1^{\prime
}+1)(2l_2^{\prime }+1)(2l+1)}{4\pi }]^{\frac 12} \\ 
\times \left( 
\begin{array}{ccc}
l_1^{\prime } & l_2^{\prime } & l \\ 
0 & 0 & 0
\end{array}
\right) \left( 
\begin{array}{ccc}
l_1 & l_2 & l \\ 
m_1 & m_2 & m
\end{array}
\right) \left\{ 
\begin{array}{ccc}
l_1 & l_2 & l \\ 
l_2^{\prime } & l_1^{\prime } & 1
\end{array}
\right\} ,
\end{array}
\eqnum{4.13}
\end{equation}
and $J_{l_il_j^{\prime }l}(k_i,k_j,q)$ is defined as the same as given in
Eq.(3.23). With the introduction of the above confining potential, the total
interaction Hamiltonian in Eq.(2.27)\ is now taken to be 
\begin{equation}
V(\alpha \beta ;\gamma \delta )=V_g(\alpha \beta ;\gamma \delta )+V_c(\alpha
\beta ;\gamma \delta ),  \eqnum{4.14}
\end{equation}
where 
\begin{equation}
V_g(\alpha \beta ;\gamma \delta )=K^0(\alpha \beta ;\gamma \delta ), 
\eqnum{4.15}
\end{equation}
which was formulated in Eqs.(3.9)-(3.24) and \ the $V_c(\alpha \beta ;\gamma
\delta )$ was given in Eqs.(4.11)-(4.13).

Now we turn to discuss the wave function $\psi (\alpha \beta )$ in Eq.(2.27)
which was defined in Eq.(2.28). In the lowest order approximation, the
two-gluon bound states can be written in the form 
\begin{equation}
|n\rangle =\sum\limits_{\alpha \beta }f_{\alpha \beta }^n{\bf a}_\alpha ^{+}%
{\bf a}_\beta ^{+}|0\rangle ,  \eqnum{4.16}
\end{equation}
where 
\begin{equation}
f_{\alpha \beta }^n=\frac 1{\sqrt{8}}\delta
_{c_1c_2}C_{l_1m_1l_2m_2}^{JM}f_{\lambda _1l_1,\lambda _2l_2}^{J\pi
}(k_1,k_2),  \eqnum{4.17}
\end{equation}
in which $\delta _{c_1c_2}$ represents the color singlet, $%
C_{l_1m_1l_2m_2}^{JM}$ is the C-G coupling coefficient, $\alpha
=(c_1,\lambda _1,l_1,m_1,k_1,\xi _\alpha =+1),$ $\beta =(c_2,\lambda
_2,l_2,m_2,k_2,\xi _\beta =+1)$, $J,M$ are the total angular momentum and
its third component of a glueball and $\pi $ denotes the spatial parity and
charge\ conjugation parity. With the introduction of the cluster coordinates 
\begin{equation}
\overrightarrow{K}=\overrightarrow{k}_1+\overrightarrow{k}_2,\ \ \ \ 
\overrightarrow{k}=\frac 12(\overrightarrow{k}_1-\overrightarrow{k}_2), 
\eqnum{4.18}
\end{equation}
where $\overrightarrow{K}$ and $\overrightarrow{k}$ are the total momentum
and relative momentum respectively, we see, in the center of mass system ($%
\overrightarrow{K}=0$), Eq.(4.17) reads 
\begin{equation}
f_{\alpha \beta }^n=\frac 1{\sqrt{8}}\delta
_{c_1c_2}C_{l_1m_1l_2m_2}^{JM}f_{\lambda _1l_1,\lambda _2l_2}^{J\pi
}(k)\delta (k_1-k_2),  \eqnum{4.19}
\end{equation}
where $k=|\overrightarrow{k}|=k_1=k_2$. Substituting Eq.(4.16) into
Eq.(2.2), we find 
\begin{equation}
\psi (\alpha \beta )=\chi _{\alpha ^{+}\beta ^{+}}(n)=\frac 1{\sqrt{8}}%
\delta _{c_1c_2}C_{l_1m_1l_2m_2}^{JM}g_{\lambda _1l_1,\lambda _2l_2}^{J\pi
h}(k)\delta (k_1-k_2),  \eqnum{4.20}
\end{equation}
where 
\begin{equation}
g_{\lambda _1l_1,\lambda _2l_2}^{J\pi h}(k)=f_{\lambda _1l_1,\lambda
_2l_2}^{J\pi }+(-1)^hf_{\lambda _2l_2,\lambda _1l_1}^{J\pi },  \eqnum{4.21}
\end{equation}
in which $h=l_1+l_2-J$. Evidently, if $\lambda _1=\lambda _2,l_1=l_2$ and $%
h=odd$, \ we have $g_{\lambda _1l_1,\lambda _2l_2}^{J\pi h}(k)=0$. This
gives a new selection rule for the glueball states. Upon substituting
Eq.(4.20) into Eq.(2.27) and noticing 
\begin{equation}
\begin{array}{l}
\alpha =(c,\lambda _1,l_1,m_1,k,+1),\beta =(c,\lambda _2,l_2,m_2,k,+1), \\ 
\rho =(c^{\prime },\lambda _3,l_3,m_3,q,+1),\sigma =(c^{\prime },\lambda
_4,l_4,m_4,q,+1)
\end{array}
\eqnum{4.22}
\end{equation}
and 
\begin{equation}
\sum\limits_\alpha \equiv \sum\limits_{c\lambda lm}\int_0^\infty dk, 
\eqnum{4.23}
\end{equation}
we finally arrive at 
\begin{equation}
\begin{array}{l}
(E-2\omega )g_{\lambda _1l_1,\lambda _2l_2}^{J\pi h}(k) \\ 
=\frac 1{8(2J+1)}\sum\limits_{\lambda _3l_3}\int_0^\infty dqV(\lambda
_1l_1k;\lambda _2l_2k;\lambda _3l_3q;\lambda _4l_4q;E)g_{\lambda
_3l_3,\lambda _4l_4}^{J\pi h}(q),
\end{array}
\eqnum{4.24}
\end{equation}
where $\omega =\sqrt{k^2+\mu ^2}$, $E$ is the mass of a glueball state given
in the center of mass frame and 
\begin{eqnarray}
&&V(\lambda _1l_1k;\lambda _2l_2k;\lambda _3l_3q;\lambda _4l_4q;E)  \nonumber
\\
&=&\sum\limits_{_{m_1m_3m_2m_4cc^{^{\prime }}M}}V(\alpha \beta ;\rho \sigma
;E)C_{l_1m_1l_2m_2}^{JM}C_{l_3m_3l_4m_4}^{JM},  \eqnum{4.25}
\end{eqnarray}
here $V(\alpha \beta ;\rho \sigma ;E)$ was given in Eq.(4.14). When the
explicit expression of $V(\alpha \beta ;\rho \sigma ;E)$ is substituted in
Eq.(4.25), one can see that the summation over $m_{1,}m_{2,}m_3,m_4$ and $M$
is easily carried out by utilizing the well-known formula for the angular
momentum coupling and the summation over the color indices $c$ and $%
c^{\prime }$ can be completed by noticing $f^{abc}f^{abc}=24$. We think, it
is unnecessary to show here the result given by these summations.

The equation in Eq.(4.24) is the eigenvalue equation used to calculate the
glueball spectrum. In the calculation, the QCD coupling constant $g$
contained in the part of Hamiltonian $V_g(\alpha \beta ;\gamma \delta )$ is
replaced by the running one which was derived in Ref.[34] recently in the
one-loop approximation and in a mass-dependent momentum space subtraction.
The coupling constant used in this calculation is\ of the form 
\begin{equation}
\alpha _s(\lambda )=\frac{\alpha _s^0}{1+\frac{\alpha _s^0}{2\pi }G(\lambda )%
},  \eqnum{4.26}
\end{equation}
where $\alpha _s(\lambda )=g^2(\lambda )/4\pi $, 
\begin{equation}
G(\lambda )=11\ln \lambda -\frac 23N_f[2+\sqrt{3}\pi -\frac 2{\lambda ^2}+(1+%
\frac 2{\lambda ^2})\frac 1\lambda \eta (\lambda )],  \eqnum{4.27}
\end{equation}
in which $N_f$ is the number of quark flavors, 
\begin{equation}
\eta (\lambda )=\sqrt{\lambda ^2-4}\ln \frac{\lambda +\sqrt{\lambda ^2-4}}2 
\eqnum{4.28}
\end{equation}
and $\lambda =\sqrt{\frac{p^2+\mu ^2}{\Lambda _{QCD}^2}}$ here $p$ is chosen
to be the transfer momentum of the exchanged gluon which may simply be taken
as $p=k-q$ for simplicity and $\Lambda _{QCD}$ is the QCD scale parameter.
The running coupling constant shown above is applicable not only in the high
energy domain, but also in the low energy regime. Particularly, in the large
momentum limit, it immediately goes over to the result obtained previously
in the minimal subtraction scheme.

\section{Numerical results and discussions}

In this section, we first show the theoretical glueball masses calculated
from the equation given in Eq.(4.24) and then make some discussions. Our
calculation is performed by using the standard program of Mathematica which
allows us to compute the effective Hamiltonian in Eq.(4.24) analytically. In
this paper, we confine ourself to investigate the low-lying glueball states
including the $0^{++},0^{-+},1^{++},1^{-+},2^{++}$ and $2^{-+}$ ground and
lower excited states whose masses are less than $4.0GeV$. Some of these
states have been investigated before in various models. We also examine the
effects of the longitudinal mode of the multipole fields and the different
sets of free parameters on the glueball masses. In our calculation, the
theoretical parameters are adjusted so as to be able to compare our results
\ to those presented recently by the lattice simulation [29]. The parameters
taken are: the gluon mass $\mu =0.42GeV$ which is comparable with $\mu
=(0.5\pm 0.2)GeV$ \ taken previously in the nonperturbative continuum
studies[38], the scale parameter $\Lambda _{QCD}=0.45GeV$ and the strength
of the confining potential $\gamma =0.18GeV^2$ , which satisfies the
relation $\mu \sim \Lambda _{QCD}$ and $\gamma \sim \Lambda _{QCD}^2$ which
make the parameters essentially depend on a single dimensional quantity.
Moreover, the value of $\gamma =0.18GeV^2$ is consistent with that of the
string tension in lattice simulations. The coupling constant $\alpha
_s^0=0.3 $ and\ quark flavor $N_f=3$. The calculated masses of glueball
states are displayed in table I. In the table, the case I and \ the case II
respectively denote the results obtained with and without considering the
contribution arising from the longitudinal mode of the multipole\ fields
which appears in the intermediate states of the matrix elements of the
interaction Hamiltonian. In the last column of the table, we quote the
results shown in Ref.[29]\ which \ were calculated by the lattice simulation.

\begin{description}
\item  Table I. The mass spectrum of two-gluon glueballs.
\end{description}

\begin{tabular}{|c|ccc|}
\hline
& \multicolumn{2}{|c}{Mass(GeV)} & \multicolumn{1}{|c|}{Mass(MeV)} \\ 
\cline{2-3}
Glueball states ($J^{PC}$) & Case I & \multicolumn{1}{|c}{Case II} & 
\multicolumn{1}{|c|}{Lattice results} \\ \hline
0$^{++}$ & 1.73 & 2.18 & \multicolumn{1}{l|}{1730(50)(80)} \\ 
& 2.66 & 3.59 & \multicolumn{1}{l|}{2670(180)(130)} \\ 
& 3.59 &  &  \\ \hline
0$^{-+}$ & 2.60 & 2.30 & 2590(40)(130) \\ 
& 3.65 & 3.78 & \multicolumn{1}{l|}{3640(60)(180)} \\ \hline
1$^{++}$ & 2.73 & 2.42 & \multicolumn{1}{l|}{} \\ 
& 3.45 & 3.51 &  \\ \hline
1$^{-+}$ & 2.67 & 2.59 & \multicolumn{1}{l|}{} \\ 
& 2.87 & 3.01 &  \\ \hline
2$^{++}$ & 2.43 & 2.43 & 2400(25)(120) \\ \hline
2$^{-+}$ & 3.32 & 2.26 & \multicolumn{1}{l|}{3100(30)(150)} \\ \hline
\end{tabular}

As seen from Eq.(4.24), each glueball state is not only assigned by its spin 
$J$ and parity $\pi ,$ but also related to the mode marked by $(\lambda
_1,l_1)(\lambda _2,l_2)$. In this paper, we take low-lying modes to perform
the calculation. For the scalar glueballs of quantum numbers $J^{PC}=0^{++}$
and the tensor ones $2^{++}$, according to the angular momentum and the
parities of the multipole fields, we take the mixture of the modes $(TE1TE1)$
and $(TM1TM1)$. For the glueball states $0^{-+}$ and $2^{-+}$, the modes are
taken to be $(TE1TM1)$ for every glueball, as was similarly done in the
investigation within the bag model [19,20]. This means that these glueballs
are mainly constructed by the gluons with transverse polarization. But, this
does not imply no contribution of the longitudinally polarized gluons to
these glueballs. The longitudinal gluons may, as virtual particles, appear
in the intermediate states in the effective interaction Hamiltonian. It is
emphasized here that for the transverse mode of gluons, as mentioned in
Appendix A, the mode $l_i=0$ is not permitted. This mode can only exist for
the longitudinal gluons. Different from the case of massless gluons, the
longitudinal mode of massive gluons is possible to take part in formation of
some glueballs. For example, the glueball states $1^{-+}$ can be formed not
only by a combination of modes $(M1E1)$ and $(E1L0)$ which gives the states
with masses as listed in the table, the modes $(M1E1)$ and $(L1L0)$ can also
form the glueball states with masses $3.23GeV$ and $3.82GeV$ respectively.
For the states $1^{++},$we only take the mode $(L0M1)$ in our calculation
because according to the B-S amplitude constructed in Sec.IV (see
Eq.(4.21)), the modes ($E1E1),(M1M1)$ are forbidden.

In order to determine the parameter dependence and errors in our
calculation, we take three sets of parameters in which we set $\mu =\Lambda
_{QCD}$ and $\gamma $ (in unit GeV$^2$)$=\Lambda _{QCD}^2$. The results of
case I in table I with different parameters (the parameters $\alpha _s^0$
and $N_f$ remain unchanged) are presented in Table II. We find that the
masses of glueballs increase gradually when the set of parameters increase.
This indicates that our numerical calculation is stable and the results are
reliable.

Table II. The mass spectrum with different parameters.

\begin{tabular}{|cccc|}
\hline
\multicolumn{1}{|c|}{} & \multicolumn{3}{|c|}{Mass(GeV)} \\ \cline{2-4}
\multicolumn{1}{|c|}{Glueball states ($J^{PC}$)} & $\mu =0.35GeV$ & $\mu
=0.45GeV$ & $\mu =0.55GeV$ \\ \hline
\multicolumn{1}{|c|}{0$^{++}$} & 1.66 & 1.74 & 1.81 \\ 
\multicolumn{1}{|c|}{} & 2.59 & 2.68 & 2.75 \\ 
\multicolumn{1}{|c|}{} & 3.52 & 3.60 & 3.66 \\ \hline
\multicolumn{1}{|c|}{0$^{-+}$} & 2.51 & 2.60 & 2.68 \\ 
\multicolumn{1}{|c|}{} & 3.57 & 3.66 & 3.73 \\ \hline
\multicolumn{1}{|c|}{1$^{++}$} & 2.64 & 2.73 & 2.82 \\ 
\multicolumn{1}{|c|}{} & 3.37 & 3.46 & 3.54 \\ \hline
\multicolumn{1}{|c|}{1$^{-+}$} & 2.59 & 2.69 & 2.78 \\ 
\multicolumn{1}{|c|}{} & 2.78 & 2.88 & 2.96 \\ \hline
\multicolumn{1}{|c|}{2$^{++}$} & 2.35 & 2.45 & 2.53 \\ \hline
\multicolumn{1}{|c|}{2$^{-+}$} & 3.25 & 3.34 & 3.41 \\ \hline
\end{tabular}

The calculated results show that the gluon mass give\ an appreciable effect
on the glueball masses. In our calculation, the mass of gluon should be
around 0.45 GeV. This fact indicates the reasonability of the QCD with
massive gluons which\ is chosen to be the starting point in our calculation.
For\ this kind of QCD, it is necessary to take the longitudinal mode of
gluons into account. As shown in Table I, when the longitudinal mode is
considered in the intermediate\ states, the theoretical masses for the
states mentioned above would be greatly improved. Otherwise, there would
occur a considerable discrepancy between the results given by this paper and
the lattice calculation. In addition, as illustrated before, the
longitudinal mode allows us to investigate more glueball states which
possibly exist in the world.

In comparison with the previous theoretical glueball masses obtained from
the B-S equation and the Dirac equation, our results give more support to
the lattice\ predictions [29-31] which are believed to be more reliable
because the lattice calculation is based on the QCD first principle and
essentially nonperturbative. Within the statistical errors existing in the
lattice calculations, our results shown in the first column of Table\ I can
be considered to be consistent with lattice predictions for the low-lying
glueballs with masses less than $3.5GeV$, especially, for the lowest scalar
glueball state $0^{++}$ with mass about $1.7GeV$\ and the tensor glueball
state $2^{++}$ with mass about $2.4GeV$.\ The achievement of the better
consistence is obviously attributed to the fact that our calculation is
fully relativistic and is\ able to include the contributions arising from
the retardation effect and longitudinal mode of the gluon field which could
not be considered in the previous investigations [23-25]. Now let us analyze
our results in some more detail. It is mentioned that the lowest scalar
glueball state $0^{++}$ and the tensor one $2^{++\text{ }}$ have been
investigated in many models and the theoretical masses are almost the same
even though there are a little difference between different calculations
[29-31]. For example, the mass of the lowest state $0^{++}$ was given by $%
1754\pm 65\pm 86MeV$ in a recent lattice calculation [31] which is different
from that given in Ref.[29]. It is expected that these states may be
identified with the pure glueball states and searched out first in future
experiments. Aside from the two states mentioned above, the\ first radial
excited state $0^{++}$ with mass $2.66GeV$\ should correspond to the state $%
2670(180)(130)MeV$ given in the lattice simulation even though whether the
latter is a pure glueball or not is still in question [32,33]. The next
radial excited state $0^{++}$ with mass $3.59GeV$ is a new prediction given
in this paper which was not predicted in the lattice simulation and\ the
other calculations. As for the pseudoscalar states $0^{-+}$ and pseudotensor
state $2^{-+}$ are all comparable with the corresponding states presented in
the lattice calculation. But, the mass of the state $2^{-+}$ is little
higher than the lattice one. In addition, we note that the states $1^{-+}$
and $1^{++}$ were not predicted in the lattice simulation, but the states $%
1^{-+}$ with masses $2.67GeV$ and the state $1^{++\text{ }}$with mass $%
2.73GeV$ are compatible with\ the recent calculation by the nonrelativistic
potential model [18].

In conclusion,\ it is emphasized that different from some previous
investigations, the calculation in this paper is based on the rigorous
three-dimensional relativistic equation satisfied by the two-gluon glueball
states which is derived from the QCD with massive gluons and represented in
the angular momentum representation. Especially, the interaction Hamiltonian
in the equation is given a complete expression which provides a firm basis
for further study. In this paper, even though we work in the relativistic
potential model with introducing phenomenologically a confining potential,
the new consideration of the retardation effect and the longitudinal mode of
the gluon fields allows us to get the improved theoretical results which are
well consistent with the lattice predictions. The only uncertainty in our
calculation arises from the introduction of confining potential. Certainly,
if a sophisticated confining potential could be found from the exact
interaction Hamiltonian derived in this and former papers, it would be
anticipated that a relativistic calculation may give more accurate
theoretical predictions.

\section{Acknowledgment}

This work was supported in part by National Natural Science Foundation of
China.

\section{Appendix A: The expression of gluon three-line vertex}

This appendix is used to derive the explicit expression of\ the gluon
three-line vertex in the angular momentum representation. As\ shown in
Eqs.(4.7), (4.29) and (4.30) in paper I, the gluon three-line vertex in the
interaction Hamiltonian is 
\begin{equation}
H_g^3=-\frac g2f^{abc}\int d^3x(\overrightarrow{A^a}\times \overrightarrow{%
A^b})\cdot (\nabla \times \overrightarrow{A^c}).  \eqnum{A.1}
\end{equation}
It is represented in the angular momentum representation as follows 
\begin{equation}
H_g^3=\sum\limits_{\alpha _1\alpha _2\alpha _3}A(\alpha _1\alpha _2\alpha
_3):{\bf a}_{\alpha _1}{\bf a}_{\alpha _2}{\bf a}_{\alpha _3}:,  \eqnum{A.2}
\end{equation}
where 
\begin{equation}
A(\alpha _1\alpha _2\alpha _3)=-\frac g2f^{abc}\int d^3x(\overrightarrow{A}%
_{\beta _1}^{\lambda _1}\times \overrightarrow{A}_{\beta _2}^{\lambda
_2})\cdot (\nabla \times \overrightarrow{A}_{\beta _3}^{\lambda _3}). 
\eqnum{A.3}
\end{equation}
Here we have set $\alpha _i=(\lambda _i,\beta _i)$ in which $\lambda
_i=TE,TM,L$ mark the transverse electric, transverse magnetic and
longitudinal modes of \ the\ multipole fields and $\beta _i=(l_i,m_i,k_i,\xi
_i).$ From now on, we use the symbols $l_im_i$ to represent the total
angular momentum. For later convenience, the relations between the multipole
fields which were mentioned in paper I are represented in the matrix form 
\begin{equation}
\nabla \times \left( 
\begin{array}{c}
\overrightarrow{A}_{lm}^{TE}(k\overrightarrow{x}) \\ 
\overrightarrow{A}_{lm}^{TM}(k\overrightarrow{x}) \\ 
\overrightarrow{A}_{lm}^L(k\overrightarrow{x})
\end{array}
\right) =k\left( 
\begin{array}{ccc}
0 & 1 & 0 \\ 
1 & 0 & 0 \\ 
0 & 0 & 0
\end{array}
\right) \left( 
\begin{array}{c}
\overrightarrow{A}_{lm}^{TE}(k\overrightarrow{x}) \\ 
\overrightarrow{A}_{lm}^{TM}(k\overrightarrow{x}) \\ 
\overrightarrow{A}_{lm}^L(k\overrightarrow{x})
\end{array}
\right) .  \eqnum{A.4}
\end{equation}
When we define 
\begin{equation}
T=\left( 
\begin{array}{ccc}
0 & 1 & 0 \\ 
1 & 0 & 0 \\ 
0 & 0 & 0
\end{array}
\right) ,  \eqnum{A.5}
\end{equation}
(A.4) can be rewritten as 
\begin{equation}
\nabla \times \overrightarrow{A}_{lm}^{\lambda _i}(k\overrightarrow{x}%
)=kT_{\lambda _i\lambda _j}\overrightarrow{A}_{lm}^{\lambda j}(k%
\overrightarrow{x}).  \eqnum{A.6}
\end{equation}
Thus, (A.3) reads 
\begin{equation}
A(\alpha _1\alpha _2\alpha _3)=-\frac g2f^{abc}k_3T_{\lambda _3\lambda
_j}\int d^3x(\overrightarrow{A}_{\beta _1}^{\lambda _1}\times 
\overrightarrow{A}_{\beta _2}^{\lambda _2})\cdot \overrightarrow{A}_{\beta
_3}^{\lambda _j}.  \eqnum{A.7}
\end{equation}

With the definition 
\begin{equation}
\overrightarrow{y}_{lm}^{(\tau )}(k\overrightarrow{x})=\sqrt{\frac 2\pi }%
kj_{l+\tau }(kr)\overrightarrow{Y}_{l,l+\tau ,m}(\widehat{x})\equiv 
\overrightarrow{y}_{ll^{\prime }m}(k\overrightarrow{x}),  \eqnum{A.8}
\end{equation}
where $l^{\prime }=l+\tau $ and $\tau =0,\pm 1$, the relations shown in
Eqs.(3.2)-(3.4) in paper I for the multipole fields can also be written in
the matrix form 
\begin{equation}
\left( 
\begin{array}{c}
\overrightarrow{A}_{lm}^{TE}(k\overrightarrow{x}) \\ 
\overrightarrow{A}_{lm}^{TM}(k\overrightarrow{x}) \\ 
\overrightarrow{A}_{lm}^L(k\overrightarrow{x})
\end{array}
\right) =\left( 
\begin{array}{ccc}
i\sqrt{\frac{l+1}{2l+1}} & 0 & -i\sqrt{\frac l{2l+1}} \\ 
0 & 1 & 0 \\ 
-i\sqrt{\frac l{2l+1}} & 0 & -i\sqrt{\frac{l+1}{2l+1}}
\end{array}
\right) \left( 
\begin{array}{c}
\overrightarrow{y}_{lm}^{(-1)}(k\overrightarrow{x}) \\ 
\overrightarrow{y}_{lm}^{(0)}(k\overrightarrow{x}) \\ 
\overrightarrow{y}_{lm}^{(1)}(k\overrightarrow{x})
\end{array}
\right) .  \eqnum{A.9}
\end{equation}
If we define $(TE,TM,L)=(-1,0,1)$ and 
\begin{equation}
B(l)=\left( 
\begin{array}{ccc}
i\sqrt{\frac{l+1}{2l+1}} & 0 & -i\sqrt{\frac l{2l+1}} \\ 
0 & 1 & 0 \\ 
-i\sqrt{\frac l{2l+1}} & 0 & -i\sqrt{\frac{l+1}{2l+1}}
\end{array}
\right) ,  \eqnum{A.10}
\end{equation}
where $l\neq 0$ and 
\begin{equation}
B(0)=\left( 
\begin{array}{ccc}
0 & 0 & 0 \\ 
0 & 0 & 0 \\ 
0 & 0 & -i
\end{array}
\right)  \eqnum{A.11}
\end{equation}
which means only the longitudinal mode survives when $l=0$, then (A.9) can
concisely be written as 
\begin{equation}
\overrightarrow{A}_{lm}^\lambda (k\overrightarrow{x})=B(l)_{\lambda \tau }%
\overrightarrow{y}_{lm}^{(\tau )}(k\overrightarrow{x}).  \eqnum{A.12}
\end{equation}
After inserting (A.12) into (A.7) and noticing the definition given in
Eq.(4.16) in paper I, we have 
\begin{equation}
\begin{array}{l}
A(\alpha _1\alpha _2\alpha _3) \\ 
=-\frac g2f^{abc}k_3\frac 1{\prod\limits_{i=1}^3\sqrt{2\omega (k_i)}}%
T_{\lambda _3\lambda _j}B^{\xi _1}(l_1)_{\lambda _1\tau _1}B^{\xi
_2}(l_2)_{\lambda _2\tau _2}B^{\xi _3}(l_3)_{\lambda _j\tau _3} \\ 
\times \int d^3x[\overrightarrow{y}_{l_1l_1^{\prime }m_1}^{\xi _1}(k_1%
\overrightarrow{x})\times \overrightarrow{y}_{l_2l_2^{\prime }m_2}^{\xi
_2}(k_2\overrightarrow{x})]\cdot \overrightarrow{y}_{l_3l_3^{\prime
}m_3}^{\xi _3}(k_3\overrightarrow{x}),
\end{array}
\eqnum{A.13}
\end{equation}
where we have defined 
\begin{equation}
\overrightarrow{y}_{ll^{\prime }m}^\xi (k\overrightarrow{x})=\left\{ 
\begin{array}{c}
\overrightarrow{y}_{ll^{\prime }m}(k\overrightarrow{x})\text{ \ if }\xi =1
\\ 
\text{\ }\overrightarrow{y}_{ll^{\prime }m}^{*}(k\overrightarrow{x})\text{ \
if }\xi =-1
\end{array}
\right.  \eqnum{A.14}
\end{equation}
and 
\begin{equation}
B^\xi (l_i)_{\lambda \tau }=\left\{ 
\begin{array}{c}
B(l_i)_{\lambda \tau }\text{ \ if }\xi =1 \\ 
\text{ \ }B^{*}(l_i)_{\lambda \tau }\text{\ if }\xi =-1
\end{array}
\right. .  \eqnum{A.15}
\end{equation}
In light of the the expression in (A.8), the integral over $\overrightarrow{x%
}$ in (A.13) can be represented as 
\begin{equation}
\begin{array}{l}
\int d^3x[\overrightarrow{y}_{l_1l_1^{\prime }m_1}^{\xi _1}(k_1%
\overrightarrow{x})\times \overrightarrow{y}_{l_2l_2^{\prime }m_2}^{\xi
_2}(k_2\overrightarrow{x})]\cdot \overrightarrow{y}_{l_3l_3^{\prime
}m_3}^{\xi _3}(k_3\overrightarrow{x}) \\ 
=(\frac 2\pi )^{\frac 32}k_1k_2k_3J_{l_1^{\prime }l_2^{\prime }l_3^{\prime
}}(k_1,k_2,k_3)\Gamma (l_i,l_i^{\prime },m_i,\xi _i),
\end{array}
\eqnum{A.16}
\end{equation}
where 
\begin{equation}
\Gamma (l_i,l_i^{\prime },m_i,\xi _i)=\int d\Omega (\widehat{x})[%
\overrightarrow{Y}_{l_1l_1^{\prime }m_1}^{\xi _1}(\widehat{x})\times 
\overrightarrow{Y}_{l_2l_2^{\prime }m_2}^{\xi _2}(\widehat{x})]\cdot 
\overrightarrow{Y}_{l_3l_3^{\prime }m_3}^{\xi _3}(\widehat{x})  \eqnum{A.17}
\end{equation}
and 
\begin{equation}
J_{l_1^{\prime }l_2^{\prime }l_3^{\prime }}(k_1,k_2,k_3)=\int
drr^2j_{l_1^{\prime }}(k_1r)j_{l_2^{\prime }}(k_2r)j_{l_3^{\prime }}(k_3r). 
\eqnum{A.18}
\end{equation}
On substituting (A.16) in (A.13), we just give\ the formula denoted in
Eq.(3.22).

First, let us calculate the $\Gamma (l_i,l_i^{\prime },m_i,\xi _i)$ in the
case of $\xi _i=1$. In this case, we set 
\begin{equation}
\Gamma (l_i,l_i^{\prime },m_i)=\int d\Omega (\widehat{x})[\overrightarrow{Y}%
_{l_1l_1^{\prime }m_1}(\widehat{x})\times \overrightarrow{Y}_{l_2l_2^{\prime
}m_2}(\widehat{x})]\cdot \overrightarrow{Y}_{l_3l_3^{\prime }m_3}(\widehat{x}%
).  \eqnum{A.19}
\end{equation}
By using the following formulas [39,40]: 
\begin{equation}
\begin{array}{l}
\overrightarrow{Y}_{ll^{\prime }m}(\widehat{x})=\sum\limits_{m^{\prime
}q}C_{l^{\prime }m^{\prime }1q}^{lm}Y_{l^{\prime }m^{\prime }}(\widehat{x})%
\overrightarrow{e}_q, \\ 
\overrightarrow{e}_{q_1}\times \overrightarrow{e}_{q_2}=i\sqrt{2}%
\sum\limits_sC_{1q_11q_2}^{1s}\overrightarrow{e}_s, \\ 
\overrightarrow{e}_q=(-1)^q\overrightarrow{e}_{-q}^{*},\text{ \ \ }%
\overrightarrow{e}_q^{*}\cdot \overrightarrow{e}_{q^{\prime }}=\delta
_{qq^{\prime }},
\end{array}
\eqnum{A.20}
\end{equation}
we find 
\begin{equation}
\begin{array}{l}
\Gamma (l_i,l_i^{\prime },m_i)=i\sqrt{2}\sum%
\limits_{n_1n_2n_3q_1q_2q_3}(-1)^{q_3}C_{l_1^{\prime
}n_11q_1}^{l_1m_1}C_{l_2^{\prime }n_21q_2}^{l_2m_2}C_{l_3^{\prime
}n_31q_3}^{l_3m_3}C_{1q_11q_2}^{1,-q_3} \\ 
\times \int d\Omega (\widehat{x})Y_{l_1^{\prime }n_1}(\widehat{x}%
)Y_{l_2^{\prime }n_2}(\widehat{x})Y_{l_3^{\prime }n_3}(\widehat{x}).
\end{array}
\eqnum{A.21}
\end{equation}
Employing the familiar formula for the above integral and the definition and
property of\ 3-j and 9-j symbols for the angular momentum couplings [39,40],
one can get 
\begin{equation}
\begin{array}{c}
\Gamma (l_i,l_i^{\prime },m_i)=i[\frac{(2l_1+1)(2l_2+1)(2l_3+1)(2l_1^{\prime
}+1)(2l_2^{\prime }+1)(2l_3^{\prime }+1)}{2\pi }]^{\frac 12} \\ 
\times \left( 
\begin{array}{ccc}
l_1^{\prime } & l_2^{\prime } & l_3^{\prime } \\ 
0 & 0 & 0
\end{array}
\right) \left( 
\begin{array}{ccc}
l_1 & l_2 & l_3 \\ 
m_1 & m_2 & m_3
\end{array}
\right) \left\{ 
\begin{array}{ccc}
1 & 1 & 1 \\ 
l_1^{\prime } & l_2^{\prime } & l_3^{\prime } \\ 
l_1 & l_2 & l_3
\end{array}
\right\} .
\end{array}
\eqnum{A.22}
\end{equation}

In the case of $\xi =-1$, it is easy to find 
\begin{equation}
\overrightarrow{Y}_{ll^{\prime }m}^{*}(\widehat{x})=(-1)^{l+l^{\prime }+m+1}%
\overrightarrow{Y}_{ll^{\prime }-m}(\widehat{x}).  \eqnum{A.23}
\end{equation}
If we define 
\begin{equation}
\overrightarrow{Y}_{ll^{\prime }m}^\eta (\widehat{x})=\left\{ 
\begin{array}{c}
\overrightarrow{Y}_{ll^{\prime }m}(\widehat{x})\text{ \ if }\xi =1 \\ 
\text{ \ }\overrightarrow{Y}_{ll^{\prime }m}^{*}(\widehat{x}))\text{ if }\xi
=-1
\end{array}
\right. ,  \eqnum{A.24}
\end{equation}
the vector spherical harmonics $\overrightarrow{Y}_{ll^{\prime }m}(\widehat{x%
})$ and $\overrightarrow{Y}_{ll^{\prime }m}^{*}(\widehat{x})$ can be
represented in an unified form 
\begin{equation}
\overrightarrow{Y}_{ll^{\prime }m}^\eta (\widehat{x})=(-1)^{(l+l^{\prime
}+m+1)\sin [\frac{(1-\eta )\pi }4]}\overrightarrow{Y}_{ll^{\prime },\eta m}(%
\widehat{x}).  \eqnum{A.25}
\end{equation}
Thus, according to (A.22)- (A.25), we have 
\begin{equation}
\begin{array}{l}
\Gamma (l_i,l_i^{\prime },m_i,\eta _i)\equiv \int d\Omega (\widehat{x})[%
\overrightarrow{Y}_{l_1l_1^{\prime }m_1}^{\eta _1}(\widehat{x})\times 
\overrightarrow{Y}_{l_2l_2^{\prime }m_2}^{\eta _2}(\widehat{x})]\cdot 
\overrightarrow{Y}_{l_3l_3^{\prime }m_3}^{\eta _3}(\widehat{x}) \\ 
=\prod\limits_{i=1}^3(-1)^{(l_i+l_i^{\prime }+m_i+1)\sin [\frac{(1-\eta
_i)\pi }4]}\Gamma (l_i,l_i^{\prime },\eta _im_i),
\end{array}
\eqnum{A.26}
\end{equation}
where $\Gamma (l_i,l_i^{\prime },\eta _im_i)$, as expressed in Eq.(3.24), is
directly written out from (A.22) with replacing $m_i$ by $\eta _im_i.$

Let us turn to compute the integral in (A.18) following the method proposed
by one of the authors in this paper and his coworker in their early
publications [41]. As we know, there is a momentum conservation in the gluon
three-line vertex: $\overrightarrow{k_{1}}+\overrightarrow{\text{ }k_{2}}+%
\overrightarrow{\text{ }k_{3}}=0$ which gives a certain restriction on the
magnitudes of the three momenta. In fact, from $k_{1}^{2}=(\overrightarrow{%
\text{ }k_{2}}+\overrightarrow{\text{ }k_{3}}%
)^{2}=k_{2}^{2}+k_{3}^{2}+2k_{2}k_{3}\cos \theta _{12},$ it is seen that
when $\cos \theta _{12}=\pm 1,$ we have $k_{1}=k_{2}+k_{3}$ and $%
k_{1}=\left\vert k_{2}-k_{3}\right\vert $. This implies that only when the
conditions $k_{1}+k_{2}\geqslant k_{3},k_{2}+k_{3}\geqslant k_{1}$ and $%
k_{1}+k_{3}\geqslant k_{2}$ or

\begin{equation}
k_1+k_2-k_3\geqslant 0,k_2+k_3-k_1\geqslant 0,k_1+k_3-k_2\geqslant 0 
\eqnum{A.27}
\end{equation}
are simultaneously satisfied, the momentum conservation holds; whereas, when 
$k_1>k_2+k_3$, the momentum conservation is violated. In addition, adding
any two inequalities in (A.27) together, we find $k_i\geqslant 0$,
therefore, each $k_i$ varies from zero to infinity. In later derivations,
the following relations are useful 
\begin{equation}
\begin{array}{l}
h_l^{(1)}(x)=j_l(x)+in_l(x),\text{ }h_l^{(2)}(x)=j_l(x)-in_l(x), \\ 
j_l(-x)=(-1)^lj_l(x),\text{ \ \ \ }h_l^{(1)}(-x)=(-1)^lh_l^{(2)}(x),
\end{array}
\eqnum{A.28}
\end{equation}
where $j_l(x)$ is the spherical Bessel function, $n_l(x)$ the spherical
Neumann function, $h_l^{(1)}(x)$ and $h_l^{(2)}(x)$ are the first class
spherical Hankel function and the second class one respectively. The
asymptotic behaviors of these functions are as follows. When $x\rightarrow 0$%
, 
\begin{equation}
\begin{array}{l}
j_l(x)\longrightarrow \frac{2^ll!x^l}{(2l+1)!},\text{ \ \ \ \ \ \ }%
n_l(x)\longrightarrow \frac{-(2l-1)!!}{x^{l+1}}, \\ 
h_l^{(1)}(x)\longrightarrow \frac{-i(2l)!}{2^ll!x^{l+1}},\text{ \ \ \ \ }%
h_l^{(2)}(x)\longrightarrow \frac{i(2l-1)!!}{x^{l+1}},
\end{array}
\eqnum{A.29}
\end{equation}
and when $x\rightarrow \infty ,$ 
\begin{equation}
\begin{array}{l}
j_l(x)\longrightarrow \frac{-i}{2x}[e^{i(x-\frac{l\pi }2)}-e^{-i(x-\frac{%
l\pi }2)}], \\ 
h_l^{(1)}(x)\longrightarrow \frac{-i}xe^{i(x-\frac{l\pi }2%
)},h_l^{(2)}(x)\longrightarrow \frac ixe^{-i(x-\frac{l\pi }2)}.
\end{array}
\eqnum{A.30}
\end{equation}

First, we prove that the integral $J_{l_1^{\prime }l_2^{\prime }l_3^{\prime
}}(k_1,k_2,k_3)$ vanishes in the case of $k_1>k_2+k_3$. In this case,
considering the analytical property of the functions $j_l(x)$ and $%
h_l^{(1)}(x)$ as shown in (A.29) and (A.30), the following integral along
the contour $C$ on the upper half complex plane of $r$ as depicted in Fig.1
is zero 
\begin{equation}
\oint_Cdrr^2h_{l_1}^{(1)}(k_1r)j_{l_2}(k_2r)j_{l_3}(k_3r)=0.  \eqnum{A.31}
\end{equation}
The contour $C$ can be divided into four parts, $C=C_0+(-\infty
,0^{-})+C_1+(0^{+},+\infty )$. Clearly, the integral along the large half
circle $C_1$ vanishes when $\left| r\right| $\ tends to infinity. Thus,
noticing the relations in (A.28), $l_i\geqslant 0$ and $l_1+l_2+l_3=even$\
which is implied by the first 3-j symbol in (A.22), one can get from (A.31) 
\begin{equation}
J_{l_1^{\prime }l_2^{\prime }l_3^{\prime }}(k_1,k_2,k_3)=-\frac 12%
\int_{C_0}drr^2h_{l_1}^{(1)}(k_1r)j_{l_2}(k_2r)j_{l_3}(k_3r).  \eqnum{A.32}
\end{equation}
\ \ \ \ \ \ \ \ Substituting the series expansions 
\begin{equation}
\begin{tabular}{l}
$j_l(kr)=\frac{\sqrt{\pi }}2\sum\limits_{\mu =0}^\infty \frac{(-1)^\mu (%
\frac{kr}2)^{2\mu +l}}{\Gamma (\mu +1)\Gamma (\mu +l+\frac 32)},$ \\ 
$n_l(kr)=\frac{\sqrt{\pi }}2\sum\limits_{\mu =0}^\infty \frac{(-1)^{l+\mu
+1}(\frac{kr}2)^{2\mu -l-1}}{\Gamma (\mu +1)\Gamma (\mu -l+\frac 12)}$%
\end{tabular}
\eqnum{A.33}
\end{equation}
into the right hand side of (A.32), it is easy to find that the integral
along the circle $C_0$ around the origin also vanishes when $\left| r\right| 
$ goes to zero. Thus, we reach the following result 
\begin{equation}
J_{l_1^{\prime }l_2^{\prime }l_3^{\prime }}(k_1,k_2,k_3)=0.  \eqnum{A.34}
\end{equation}

Next, we compute the integral under the conditions shown in (A.27). In view
of these conditions and the asymptotic behaviors of $h_l^{(1)}(x)$ and \ $%
h_l^{(2)}(x)$ shown in (A.30), the function $f(r)$ defined by 
\begin{equation}
\begin{array}{l}
f(r)=h_{l_1}^{(1)}(k_1r)h_{l_2}^{(1)}(k_2r)h_{l_3}^{(2)}(k_3r)+h_{l_1}^{(1)}(k_1r)h_{l_2}^{(2)}(k_2r)h_{l_3}^{(1)}(k_3r)
\\ 
+h_{l_1}^{(2)}(k_1r)h_{l_2}^{(1)}(k_2r)h_{l_3}^{(1)}(k_3r)+h_{l_1}^{(1)}(k_1r)h_{l_2}^{(1)}(k_2r)h_{l_3}^{(1)}(k_3r)
\end{array}
\eqnum{A.35}
\end{equation}
is analytical on the upper half complex plane of $r$ except for at the
origin. Therefore, we have 
\begin{equation}
\int_Cdrr^2f(r)=0,  \eqnum{A.36}
\end{equation}
where the contour $C$ is still represented in Fig.1. Due to the conditions
in (A.27), the integral along$\;C_1$ still vanishes. Thus,\ from the above
integral, we get 
\begin{equation}
\begin{tabular}{l}
$J_{l_1^{\prime }l_2^{\prime }l_3^{\prime }}(k_1,k_2,k_3)=\frac 18%
\{\int_0^\infty drr^2f(r)+\int_{-\infty }^0drr^2f(r)\}$ \\ 
$=-\frac 18\int_{C_0}drr^2f(r).$%
\end{tabular}
\eqnum{A.37}
\end{equation}
In accordance with (A.28), the function $f(r)$ can be written as 
\begin{equation}
\begin{tabular}{l}
$f(r)$ \\ 
$%
=4j_{l_1}(k_1r)j_{l_2}(k_2r)j_{l_3}(k_3r)+2in_{l_1}(k_1r)j_{l_2}(k_2r)j_{l_3}(k_3r) 
$ \\ 
$%
+2ij_{l_1}(k_1r)n_{l_2}(k_2r)j_{l_3}(k_3r)+2ij_{l_1}(k_1r)j_{l_2}(k_2r)n_{l_3}(k_3r) 
$ \\ 
$+2in_{l_1}(k_1r)n_{l_2}(k_2r)n_{l_3}(k_3r).$%
\end{tabular}
\eqnum{A.38}
\end{equation}
Inserting this expression into (A.37) and using the series representation in
(A.33), it can be found that\ except for the last term, the other terms in
(A.38) all give no contribution to the integral. Therefore, we have 
\begin{equation}
\begin{array}{l}
J_{l_1^{\prime }l_2^{\prime }l_3^{\prime }}(k_1,k_2,k_3)=-\frac i4%
\int_{C_0}drr^2n_{l_1}(k_1r)n_{l_2}(k_2r)n_{l_3}(k_3r) \\ 
=\frac{i\pi ^{\frac 32}}{32}\sum\limits_{\mu _1,\mu _2,\mu
_3}\prod\limits_{i=1}^3\frac{(-1)^{\mu _i+l_i}(\frac{k_i}2)^{2\mu _i-l_i-1}}{%
\Gamma (\mu _i+1)\Gamma (\mu _i-l_i+\frac 12)}\int_{C_0}drr^{2(\mu _1+\mu
_2+\mu _3)-l_1-l_2-l_3-1}.
\end{array}
\eqnum{A.39}
\end{equation}
Setting $r=\rho e^{i\theta }$ and noticing $2(\mu _1+\mu _2+\mu
_3)-l_1-l_2-l_3=even$ and $\int_\pi ^0d\theta e^{i2m\theta }=-\pi \delta
_{m,0}$ here $m$ is an integer, we finally obtain the expression as shown in
Eq.(3.23).

\section{Appendix B: The expression of gluon four-line vertex}

In this appendix we would like to derive the explicit expression of the
gluon four-line vertex in the angular momentum representation for
completeness although the vertex gives no contribution to the equation
(2.27) in the lowest order approximation due to $S_2(\rho \sigma \tau
\lambda ,\gamma \delta )=0$ as shown in Eq.(3.1). The four-line vertex in
the interaction Hamiltonian\ at $t=0$ which was described in Eqs.(4.7),
(4.29) and (4.31) in paper I may be rewritten as 
\begin{equation}
\begin{tabular}{l}
$H_g^4=\frac{g^2}4f^{abe}f^{cde}\int d^3x(\overrightarrow{A^a}\cdot 
\overrightarrow{A^c})(\overrightarrow{A^b}\cdot \overrightarrow{A^d})$ \\ 
$=\sum\limits_{\alpha \beta \gamma \delta }B(\alpha _1\alpha _2\alpha
_3\alpha _4):{\bf a}_{\alpha _1}{\bf a}_{\alpha _2}{\bf a}_{\alpha _3}{\bf a}%
_{\alpha _4}:,$%
\end{tabular}
\eqnum{B.1}
\end{equation}
where the second equality is obtained by substituting the expansion of gluon
fields in terms of the multipole fields into the first equality and the
coefficient function is of the form 
\begin{equation}
B(\alpha _1\alpha _2\alpha _3\alpha _4)=\frac{g^2}4f^{abe}f^{cde}\int d^3x(%
\overrightarrow{A}_{\alpha _1}^{\lambda _1}\cdot \overrightarrow{A}_{\alpha
_2}^{\lambda _2})(\overrightarrow{A}_{\alpha _3}^{\lambda _3}\cdot 
\overrightarrow{A}_{\alpha _4}^{\lambda _4}).  \eqnum{B.2}
\end{equation}
By making use of the representation in (A.8)-(A.12), (A.14) and (A.15) and
noticing the definition given in Eq.(4.16) in paper I, the $B(\alpha
_1\alpha _2\alpha _3\alpha _4)$ can be represented as 
\begin{equation}
\begin{tabular}{l}
$B(\alpha _1\alpha _2\alpha _3\alpha _4)=\frac{g^2}{\pi ^2}%
f^{abe}f^{cde}\prod\limits_{i=1}^4\frac{k_i}{\sqrt{2\omega (k_i)}}B^{\xi
_i}(l_i)_{\lambda _i\tau _i}$ \\ 
$\times J_{l_1^{^{\prime }}l_2^{^{\prime }}l_3^{^{\prime }}l_4^{\prime
}}(k_1,k_2,k_3,k_4)\widetilde{\Gamma }(l_i,l_i^{\prime },m_i,\eta _i),$%
\end{tabular}
\eqnum{B.3}
\end{equation}
where 
\begin{equation}
J_{l_1^{\prime }l_2^{\prime }l_3^{\prime }l_4}(k_1,k_2,k_3,k_4)=\int
drr^2j_{l_1^{\prime }}(k_1r)j_{l_2^{\prime }}(k_2r)j_{l_3^{\prime
}}(k_3r)j_{l_4^{\prime }}(k_4r)  \eqnum{B.4}
\end{equation}
and 
\begin{equation}
\widetilde{\Gamma }(l_i,l_i^{\prime },m_i,\eta _i)=\int d\Omega (\widehat{x})%
\overrightarrow{Y}_{l_1l_1^{\prime }m_1}^{\eta _1}(\widehat{x})\cdot 
\overrightarrow{Y}_{l_2l_2^{\prime }m_2}^{\eta _2}(\widehat{x})%
\overrightarrow{Y}_{l_3l_3^{\prime }m_3}^{\eta _3}(\widehat{x})\cdot 
\overrightarrow{Y}_{l_4l_4^{\prime }m_4}^{\eta _4}(\widehat{x}),  \eqnum{B.5}
\end{equation}
here the notation in (A.24) has been used. Inserting (A.20) and (A.25) into
(B.5) and employing the familiar formulas for the integrals of spherical
harmonics and for the angular momentum coupling [39,40], it is not difficult
to get 
\begin{equation}
\begin{tabular}{l}
$\widetilde{\Gamma }(l_i,l_i^{\prime },m_i,\eta _i)$ \\ 
$=\frac 1{4\pi }\sum_{lm}(-1)^{l_1+l_2+l_3^{\prime }+l_4^{\prime
}+m}(2l+1)\prod_{i=1}^4(-1)^{(l_i+l_i^{\prime }+m_i+1)\sin [\frac{(1-\eta
_i)\pi }4]}$ \\ 
$\times [(2l+1)(2l^{\prime }+1)]^{\frac 12}\left( 
\begin{array}{ccc}
l_1^{\prime } & l_3^{\prime } & l \\ 
0 & 0 & 0
\end{array}
\right) \left( 
\begin{array}{ccc}
l_2^{\prime } & l_4^{\prime } & l \\ 
0 & 0 & 0
\end{array}
\right) \left\{ 
\begin{array}{ccc}
l_1 & l_3 & l \\ 
l_3^{\prime } & l_1^{\prime } & 1
\end{array}
\right\} $ \\ 
$\times \left\{ 
\begin{array}{ccc}
l_2 & l_4 & l \\ 
l_4^{\prime } & l_2^{\prime } & 1
\end{array}
\right\} \left( 
\begin{array}{ccc}
l_1 & l_3 & l \\ 
\eta _1m_1 & \eta _3m_3 & m
\end{array}
\right) \left( 
\begin{array}{ccc}
l_2 & l_4 & l \\ 
\eta _2m_2 & \eta _4m_4 & -m
\end{array}
\right) .$%
\end{tabular}
\eqnum{B.6}
\end{equation}

To compute the integral in (B.4), we first examine the conditions satisfied
by the magnitudes of the momenta. From the momentum conservation $%
\overrightarrow{k_1}+\overrightarrow{k_2}+\overrightarrow{k_3}+%
\overrightarrow{k_4}=0$, we have 
\begin{equation}
\begin{tabular}{l}
$k_1^2=(\overrightarrow{k_2}+\overrightarrow{k_3}+\overrightarrow{k_4}%
)^2=k_2^2+k_3^2+k_4^2$ \\ 
$+2k_2k_3\cos \theta _{23}+2k_2k_4\cos \theta _{24}+2k_3k_4\cos \theta
_{34}. $%
\end{tabular}
\eqnum{B.7}
\end{equation}
From the above equality, we may find the maximum and minimum of $k_1$ by
setting \ $\theta _{ij}=0$ or $\pi $. There are only four ways which permit
us to take the values of $\theta _{ij}=0$ or $\pi $. These are (1) $\theta
_{23}=\theta _{24}=\theta _{34}=0$; (2)\ $\theta _{23}=\theta _{24}=\pi
,\theta _{34}=0$; (3) $\theta _{23}=\theta _{34}=\pi ,\theta _{24}=0$; (4) $%
\theta _{24}=\theta _{34}=\pi ,\theta _{23}=0$. Correspondingly, we get from
(B.7) the equalities $k_1=k_2+k_3+k_4,k_1=\left| k_2-k_3-k_4\right|
,k_1=\left| k_3-k_2-k_4\right| $ and $k_1=\left| k_4-k_2-k_3\right| .$ From
these equalities we find the following restriction conditions which are
consistent with the momentum conservation 
\begin{equation}
\begin{tabular}{l}
$k_2+k_3+k_4-k_1\geqslant 0,k_1+k_3+k_4-k_2\geqslant 0,$ \\ 
$k_1+k_2+k_4-k_3\geqslant 0,k_1+k_2+k_3-k_4\geqslant 0,$ \\ 
$k_1+k_2-k_3-k_4\geqslant 0,k_1+k_3-k_2-k_4\geqslant 0,$ \\ 
$k_1+k_4-k_2-k_3\geqslant 0,k_1+k_2+k_3+k_4\geqslant 0.$%
\end{tabular}
\eqnum{B.8}
\end{equation}
By adding some two of the above inequalities, one may see $k_i\geqslant 0.$
And similar to the proof described in the preceding appendix, it can be
proved that the integral in (B.4) is merely nonvanishing\ provided that the
conditions in (B.8) are respected. According to the above conditions, it is
obvious that the function defined below is analytical on upper half complex
plane of $r$ 
\begin{equation}
\begin{array}{l}
F(r) \\ 
=h_{l_1}^{(1)}(k_1r)h_{l_2}^{(1)}(k_2r)h_{l_3}^{(1)}(k_3r)h_{l_4}^{(2)}(k_4r)+h_{l_1}^{(1)}(k_1r)h_{l_2}^{(1)}(k_2r)h_{l_3}^{(2)}(k_3r)h_{l_4}^{(1)}(k_4r)
\\ 
+h_{l_1}^{(1)}(k_1r)h_{l_2}^{(2)}(k_2r)h_{l_3}^{(1)}(k_3r)h_{l_4}^{(1)}(k_4r)+h_{l_1}^{(2)}(k_1r)h_{l_2}^{(1)}(k_2r)h_{l_3}^{(1)}(k_3r)h_{l_4}^{(1)}(k_4r)
\\ 
+h_{l_1}^{(1)}(k_1r)h_{l_2}^{(1)}(k_2r)h_{l_3}^{(2)}(k_3r)h_{l_4}^{(2)}(k_4r)+h_{l_1}^{(1)}(k_1r)h_{l_2}^{(2)}(k_2r)h_{l_3}^{(1)}(k_3r)h_{l4}^{(2)}(k_4r)
\\ 
+h_{l_1}^{(1)}(k_1r)h_{l_2}^{(2)}(k_2r)h_{l_3}^{(2)}(k_3r)h_{l_4}^{(1)}(k_4r+h_{l_1}^{(1)}(k_1r)h_{l_2}^{(1)}(k_2r)h_{l_3}^{(1)}(k_3r)h_{l_4}^{(1)}(k_4r)
\end{array}
\eqnum{B.9}
\end{equation}
where the $l_i^{^{\prime }}$ have been replaced by $l_i$ for convenience.
Therefore, based on the Cauchy theorem, the integral along the contour $C$
as depicted in Fig.1 is zero, 
\begin{equation}
\int_Cdrr^2F(r)=0  \eqnum{B.10}
\end{equation}
From this equation, noticing \ $l_1+l_2+l_3+l_4=even$ as implied by the
first two 3-j symbols in (B.6), we obtain 
\begin{equation}
\begin{tabular}{l}
$J_{l_1^{\prime }l_2^{\prime }l_3^{\prime }l_4}(k_1,k_2,k_3,k_4)=\frac 1{16}%
\{\int_0^\infty drr^2F(r)+\int_{-\infty }^0drr^2F(r)\}$ \\ 
$=-\frac 1{16}\int_{C_0}drr^2F(r).$%
\end{tabular}
\eqnum{B.11}
\end{equation}
In view of the relations in (A.28), the function $F(r)$ can be represented
as 
\begin{equation}
\begin{tabular}{l}
$F(r)$ \\ 
$%
=8j_{l_1}(k_1r)j_{l_2}(k_2r)j_{l_3}(k_3r)j_{l_4}(k_4r)+2j_{l_1}(k_1r)j_{l_2}(k_2r)j_{l_3}(k_3r)n_{l_4}(k_4r) 
$ \\ 
$%
+2j_{l_1}(k_1r)j_{l_2}(k_2r)n_{l_3}(k_3r)j_{l_4}(k_4r)+2ij_{l_1}(k_1r)n_{l_2}(k_2r)j_{l_3}(k_3r)j_{l_4}(k_4r) 
$ \\ 
$%
+6in_{l_1}(k_1r)j_{l_2}(k_2r)j_{l_3}(k_3r)j_{l_4}(k_4r)+2in_{l_1}(k_1r)j_{l_2}(k_2r)n_{l_3}(k_3r)n_{l_4}(k_4r) 
$ \\ 
$%
+2in_{l_1}(k_1r)n_{l_2}(k_2r)j_{l_3}(k_3r)n_{l_4}(k_4r)+2in_{l_1}(k_1r)n_{l_2}(k_2r)n_{l_3}(k_3r)j_{l_4}(k_4r) 
$ \\ 
$-2ij_{l_1}(k_1r)n_{l_2}(k_2r)n_{l_3}(k_3r)n_{l_4}(k_4).$%
\end{tabular}
\eqnum{B.12}
\end{equation}

Upon inserting the above expression into (B.11), using the series
representation in (A.33) and considering the relations among the angular
momenta which are implied by the 3-j and 9-j symbols in (B.6), \ it is
easily verified that the first five terms in (B.12) give vanishing
contributions to the integral. The nonvanishing contributions given by the
last four terms can be calculated by the same procedure as described in the
last part of Appendix A. The result is 
\begin{equation}
\begin{tabular}{l}
$J_{l_1^{\prime }l_2^{\prime }l_3^{\prime }l_4}(k_1,k_2,k_3,k_4)$ \\ 
$=\frac{\pi ^3}{16}(-)^{\frac 12(l_1+l_2+l_3+l_4)}\stackrel{\infty }{%
\sum\limits_{\mu _1,\mu _2,\mu _3,\mu _4=0}}\delta _{2(\mu _1+\mu _2+\mu
_3+\mu _4),l_2^{\prime }+l_3^{\prime }+l_4^{\prime }-l_1^{\prime }}$ \\ 
$\times \sum\limits_{i=1}^4\frac{(1-2\delta _{i1})}{\Gamma (\mu _i+1)\Gamma
(\mu _i+l_i+\frac 32)}\prod\limits_{i=1(j\neq i)}^4\frac{(k_j/k_1)^{2\mu
_j-l_j}}{k_j\Gamma (\mu _j+1)\Gamma (\mu _j-l_j+\frac 12)}.$%
\end{tabular}
\eqnum{B.13}
\end{equation}

\section{References}

[1] Crystal Barrel Collaboration, Phys. Lett. B323, 223 (1994).

[2] V. V. Anisovich, et. al., Phys. Rev. D50, 1972 (1994); Phys. Lett. B364,
195 (1995).

[3] R. M. Baltrusaitis, et. al., Phys. Rev. Lett. 56, 107 (1986).

[4] J. Z. Bai, et. al., Phys. Rev. Lett. 76, 3502 (1996).

[5] R. M. Barnett, et. al., Phys. Rev. D54, 1 (1996).

[6] C. Amsler, et. al., Phys. Lett. B355, 425 (1995); Phys. Lett. B353, 385
(1995).

[7] D. Weingarten, Nucl. Phys. Proc. Suppl. B53, 232 (1997).

[8] J. Weinstein and N. Isgur, Phys. Rev. D27, 588 (1983); D41, 2236 (1990);
D43, 95 (1991).

[9] K. T. Chao, Commu. Theor. Phys. 24, 373 (1995) .

[10] J. Sexton, A. Vaccarino and D. Weigarten, Phys. Rev. Lett. 75, 4563
(1995).

[11] V. V. Anisovich, Phys. Lett. B364, 195 (1995).

[12] D. Barberis et al., Phys. Lett. B453, 305 (1999).

[13] L. Burakovsky, P.R. Page, Eur. Phys. J. C12, 489 (2000).

[14] D.V. Bugg, M. Peardon, B.S. Zou, Phys. Lett. B486, 49 (2000).

[15] Z. S. Wang, Phys. Rev. D62, 017503 (2000).

[16] T. Barnes, Z. Phys. C10, 275 (1981).

[17] J. M. Cornwall and A. Soni, Phys. Lett. B120, 431 (1983).

[18] W. S. Hou and G. G. Wong, Phys. Rev. D67, 034003 (2003).

[19] R. L. Jaffe and K. Johnson, Phys. Lett. B60, 201 (1976); Phys. Rev.
Lett. 34, 1645 1976).

[20] T. Barnes, F. E. Close and S. Monagham, Nucl. Phys. B198, 380 (1982).

[21] S. Narision, Z. Phys. C26, 209 (1984);\ S. Narision, Nucl. Phys. B509,
312 1998).

[22] A. Shpenik, Y. Fekete, J. Kis, Nucl. Phys. Proc. Suppl. A99, 274 (2001).

[23] S. Bhatnagar and A. N. Mitra, Nuovo \ Cim. A104, 925 (1991).

[24] M. H. Thoma, M. L\"ust and H. J. Mang, J. Phys. G: Nucl. Part. Phys.
18, 1125 (1992).

[25] J. Y. Cui, J. M. Wu, H. Y. Jin, Phys. Lett. B424, 381 (1998).

[26] K.G. Wilson, Phys. Rev. D10, 2445 (1974).

[27] C. Bernard, Phys. Lett. 108B, 431 (1982); Nucl. Phys. B219, 341 (1983).

[28] C. Michael and M. Teper, Nucl. Phys. Lett. B314, 347 (1989).

[29] C. Morningstar and M. Peardon, Phys. Rev. D60, 034509 (1999).

[30] C. Liu, Chin. Phys. Lett. 18, 187 (2001).

[31] D. Q. Liu, J. M. Wu, Y. Chen, High \ Ener. Phys. Nucl. Phys. 26, 222
(2002).

[32] G. B. West, talk given at the Montpelier Conference and Pairs Workshop
on QCD, hep-ph/9608258.

[33] W. Ochs, talk at EPS-HEP '99 Conference, July 15-21, 1999, Tampere,
Finland, hep-ph/9909241.

[34] J. C. Su, Commun. Theor. Phys. 36, 665 (2001).

[35] J. C. Su, Commun. Theor. Phys. 15, 229 (1991); Commun. Theor. Phys. 18,
327 (1992).

[36] J. C. Su, Y. B. Dong and S. S. Wu, J. Phys. G: Nucl. Part. Phys. 18,
1347 (1992).

[37] E. Eichten, et. al., Phys. Rev. Lett. 34, 369 (1975); Phys. Rev.
D21,511 (1980).

[38] J. M. Cornwall, Phys. Rev. D26, 1453 (1982).

[39] M. E. Rose, Multipole Fields, John Wiley \& Sons, New York, 1955.

[40] A. R. Edmonds, Angular \ Momentumin \ in \ Quantum Mechanics, Princeton
University Press, 1960.

[41] Sun Chia-Chung and Su Chan (the another name of Jun-Chen Su), SCIENTA
SINICA\ Vol. XIX, No. 1, 91 (1976); Vol. XXI, No. 3, 327 (1978).

\section{\protect\bigskip Figure caption}

Fig.1. The contour for the integrals containing three and four spherical
Bessel functions.

\end{document}